\documentclass[useAMS,usenatbib]{mn2e}

\usepackage{graphicx}
\usepackage{txfonts}
\usepackage{amssymb,alltt}
\usepackage{color}
\usepackage{hyperref}
\hypersetup{colorlinks, linkcolor=blue}

\title[HFF:  Geometry and dynamics of MACSJ0416.1$-$2403]{\textit{Hubble Frontier Fields:} The Geometry and Dynamics of the Massive Galaxy Cluster Merger MACSJ0416.1$-$2403}
\author[Jauzac et al. 2014]
{Mathilde Jauzac$^{1,2}$\thanks{E-mail: mathilde.jauzac@dur.ac.uk},
 Eric Jullo$^3$, Dominique Eckert$^{4}$, Harald Ebeling$^{5}$, Johan Richard$^{6}$,
\newauthor
Marceau Limousin$^{3}$, Hakim Atek$^{7}$, Jean-Paul Kneib$^{7,3}$, Benjamin Cl\'ement$^{8}$, Eiichi Egami$^{8}$,  
\newauthor
David Harvey$^{9}$, Kenda Knowles$^{2}$, Richard Massey$^{1}$, Priyamvada Natarajan$^{10}$, 
\newauthor
Beno\^it Neichel$^{3}$, M. Rexroth$^{7}$  
\\
\\
$^{1}$Institute for Computational Cosmology, Durham University, South Road, Durham DH1 3LE, U.K.\\
$^{2}$Astrophysics and Cosmology Research Unit, School of Mathematical Sciences, University of KwaZulu-Natal, Durban 4041, South Africa\\
$^{3}$Laboratoire d'Astrophysique de Marseille - LAM, Universit\'e d'Aix-Marseille $\&$ CNRS, UMR7326, 38 rue F. Joliot-Curie, 13388 Marseille Cedex 13, France\\
$^{4}$Astronomy Department, University of Geneva, 16 ch. d'Ecogia, CH-1290 Versoix, Switzerland \\
$^{5}$Institute for Astronomy, University of Hawaii, 2680 Woodlawn Drive, Honolulu, Hawaii 96822, USA\\
$^{6}$CRAL, Observatoire de Lyon, Universit\'e Lyon 1, 9 Avenue Ch. Andr\'e, 69561 Saint Genis Laval Cedex, France\\
$^{7}$Laboratoise d'Astrophysique, Ecole Polytechnique F\'ed\'erale de Lausanne (EPFL), Observatoire de Sauverny, CH-1290 Versoix, Switzerland\\
$^{8}$Steward Observatory, University of Arizona, 933 North Cherry Avenue, Tucson, AZ, 85721, USA \\
$^{9}$SUPA, University of Edinburgh, Royal Observatory, Blackford Hill, Edinburgh, EH9 3HJ, UK\\
$^{10}$Department of Astronomy, Yale University, 260 Whitney Avenue, New Haven, CT 06511, USA
}

\begin{document}

\date{Accepted 2013 XXX. Received 2013 XXX; in original form 2014 October 22nd}

\pagerange{\pageref{firstpage}--\pageref{lastpage}} \pubyear{2014}

\maketitle

\label{firstpage}

\begin{abstract}
We use a joint  optical/X-ray analysis to constrain the geometry and history of the ongoing merging event in the massive galaxy cluster  MACSJ0416.1$-$2403 ($z$=0.397).  Our investigation of cluster substructure rests primarily on a combined strong- and weak-lensing mass reconstruction based on the deep, high-resolution images obtained for the \textit{Hubble Frontier Fields} initiative. To reveal the system's dynamics, we complement this lensing analysis with a study of the intra-cluster gas using shallow {\sl Chandra} data, and a three-dimensional model of the distribution and motions of cluster galaxies derived from over 100 spectroscopic redshifts.  The multi-scale grid model obtained from our combined lensing analysis extends the high-precision strong-lensing mass reconstruction recently performed to cluster-centric distances of almost 1 Mpc.
Our analysis detects the two well known mass concentrations in the cluster core.  A pronounced offset between collisional and collisionless matter is only observed for the SW cluster component, while excellent alignment is found for the NE cluster.  Both the lensing analysis and the distribution of cluster light strongly suggest the presence of a third massive structure, almost 2 arcmin SW of the cluster centre.  Since no X-ray emission is detected in this region, we conclude that this structure is non-virialised  and speculate that it might be part of a large-scale filament almost aligned with our line of sight.  Combining all evidence from the distribution of dark and luminous matter, we propose two alternative scenarios for the trajectories of the components of MACSJ0416.1$-$2403.  Upcoming deep X-ray observations that allow the detection of shock fronts, cold cores, and sloshing gas (all key diagnostics for studies of cluster collisions) will allow us to test, and distinguish between these two scenarios.

\end{abstract}



\section{Introduction}
In the course of the past decades, gravitational lensing has become one of the most powerful tools to map the distribution of dark matter, starting with the confirmation of gravitational lensing as the causal origin of the giant arc in the cluster  Abell 370 by \cite{soucail88}. The bending of light by foreground clusters can be observed in two regimes: the strong-lensing regime, limited to the densest part of the cluster, i.e. its core, and the weak-lensing regime, in its outskirts.
Gravitational lensing allows astronomers not only to directly measure the distribution of the total gravitational mass (dark or luminous), but also to use clusters as `cosmic telescopes' to image very distant galaxies and to constrain the geometry of the Universe \citep[for reviews, see e.g. ][]{massey10,KN11}.  As the most powerful gravitational telescopes, massive clusters are sought-after observational targets.  

Occupying the nodes of the  \emph{Cosmic Web} of large-scale filaments and sheets \citep[][]{bond96}, massive clusters and, specifically, massive cluster mergers are of particular interest also in the context of structure-formation studies. 
The case of the \textit{Bullet Cluster} \citep[1E0657$-$56, z$=$0.3,][]{clowe04} is exceptional in this context: it shows a merging event of two clusters where the merging direction is perpendicular to the line of sight, maximizing the apparent separation and revealing a clear cone-shaped shock front ahead of the smaller merger component. This rare geometry has allowed studies to separately investigate the distribution and dynamics of the baryonic and dark matter components, e.g., by using a combination of strong- and weak-gravitational lensing \citep{bradac06}. This approach is particularly powerful as strong lensing constrains precisely the location and shape of the cluster core, while weak lensing maps the mass distribution on larger scales.  A similar analysis was performed also on MACSJ0025.4-1222 \citep[$z=0.58$,][]{ebeling07,bradac08b}.
 Although any given observation captures no more than a snapshot of the complex process of cluster growth, the different dynamical behaviour of collisional (gas) and collisionless matter (dark matter and galaxies) often observed in merging systems has been used to great effect to constrain the three-dimensional trajectories of the merger components \citep[e.g.,][]{clowe04,ma09,merten11,jauzac12,hsu13}.
In this paper, we investigate a massive cluster that is both a spectacularly efficient gravitational lens and an active merger.

MACSJ0416.1$-$2403 ($z=0.397$; hereafter MACSJ0416) was discovered by the Massive Cluster Survey \citep[MACS; ][]{ebeling01} and classified as an actively merging system by \citet[][]{ME12} based on its X-ray / optical morphology. Because of its large Einstein radius, as revealed in HST (Hubble Space Telescope) observations obtained for programme GO-11103 (PI: Ebeling), MACSJ0416 was selected as one of five ``high magnification'' clusters in the Cluster Lensing And Supernova survey with Hubble \citep[CLASH: ][]{postman12}. The system's highly elongated mass distribution, typical of merging systems, allowed numerous strongly lensed galaxies to be discovered \citep[][]{zitrin13a,richard14} in these imaging data.  More recently, MACSJ0416 was chosen as one of six targets for the \emph{Hubble Frontier Fields} (HFF) initiative. Launched by the Space Telescope Science Institute in 2013, this observing programme aims to harness the gravitational magnification of massive cluster lenses to study the distant Universe to unprecedented depth. The HFF programme allocates 140 HST orbits to imaging observations of each cluster, split between three filters on the \emph{Advanced Camera for Survey} (ACS), and four on the \emph{Wide Field Camera 3} (WFC3), to reach an unprecedented depth for cluster studies of $m_{\rm AB} \sim 29$ in all 7 passbands. The HFF observations of MACSJ0416 with ACS, performed in early 2014, allowed us to identify 51 new multiply imaged galaxies, bringing the total number of lensed images to a record of 194 \citep[][]{jauzac14}. The resulting strong-lensing mass model confirmed the bimodal mass distribution of MACSJ0416, and constrains the mass within the core region to a precision of better than 1\%.

In this paper, we extend the analysis of \cite{jauzac14} by using both  strong- and weak-gravitational-lensing constraints to measure and map the mass distribution of MACSJ0416 to larger cluster-centric radii. In addition we use archival \emph{Chandra} X-ray data as well as radial velocities measured for over 100 cluster galaxies to investigate the relative motions of collisional and collisionless matter in projection, as well as along our line of sight. The result is a model of the three-dimensional geometry and merger history of this complex system.
Our paper is organised as follows: observations of MACSJ0416 are summarised in Section 2, an overview of our earlier strong-lensing analysis is provided in Section 3, the construction of the weak-lensing catalogue is described in Section 4, our gravitational lensing mass-modeling technique is explained in Section 5, results are presented in Section 6, the dynamical analysis of MACSJ0416 is performed in Section 7, and, finally, a summary is provided in Section 8.
We adopt the $\Lambda$CDM concordance cosmology with $\Omega_m = 0.3$, $\Omega_{\Lambda} = 0.7$, and a $Hubble$ constant $H_0 = 70$ km s$^{-1}$ Mpc$^{-1}$. Magnitudes are quoted in the AB system.

\begin{table*}
\begin{center}
\begin{tabular}{ccccc}
\hline
\hline
R.A. (J2000) & Dec. (J2000) & Instrument/Filter & Exposure Time (in sec.) & Programme ID \\
\hline
04 16 07.2 & -24 03 35.7 & WFC2/F814W & 1200 & 11103 \\
04 16 07.2 & -24 03 35.7 & WFC2/F606W & 1200 & 11103 \\
\hline
04 16 08.4 & -24 04 20.0 & ACS/F435W & 2052 & 12459 \\
04 16 08.4 & -24 04 20.0 & ACS/F475W & 2064 & 12459 \\
04 16 08.4 & -24 04 20.0 & ACS/F606W & 2018 & 12459 \\
04 16 08.4 & -24 04 20.0 & ACS/F625W & 2017 & 12459 \\
04 16 08.4 & -24 04 20.0 & ACS/F775W & 2031 & 12459 \\
04 16 08.4 & -24 04 20.0 & ACS/F814W & 4037 & 12459 \\
04 16 08.4 & -24 04 20.0 & ACS/F850LP & 4086 & 12459 \\
04 16 08.4 & -24 04 20.0 & WFC3/F105W & 2815 & 12459 \\
04 16 08.4 & -24 04 20.0 & WFC3/F110W & 2515 & 12459 \\
04 16 08.4 & -24 04 20.0 & WFC3/F125W & 2515 & 12459 \\
04 16 08.4 & -24 04 20.0 & WFC3/F140W & 2312 & 12459 \\
04 16 08.4 & -24 04 20.0 & WFC3/F160W & 5029 & 12459 \\
04 16 08.4 & -24 04 20.0 & WFC3/F225W & 3634 & 12459 \\
04 16 08.4 & -24 04 20.0 & WFC3/F275W & 3684 & 12459 \\
04 16 08.4 & -24 04 20.0 & WFC3/F336W & 2360 & 12459 \\
04 16 08.4 & -24 04 20.0 & WFC3/F390W & 2407 & 12459 \\
\hline
\textbf{04 16 08.9} & \textbf{-24 04 28.7} & \textbf{ACS/F435W} & \textbf{52460} & \textbf{HFF-13496} \\
\textbf{04 16 08.9} & \textbf{-24 04 28.7} & \textbf{ACS/F606W} & \textbf{31476} & \textbf{HFF-13496} \\
\textbf{04 16 08.9} & \textbf{-24 04 28.7} & \textbf{ACS/F814W} & \textbf{125904} & \textbf{HFF-13496} \\
\hline
\hline
\smallskip
\end{tabular}
\end{center}
\caption{Summary of HST observations of MACSJ0416. The HFF observations are highlighted in bold.}
\label{HFFdata}
\end{table*}

\section{Observations}

\subsection{Pre-HFF HST Data}
 MACSJ0416 was first  observed with the \textit{Hubble Space Telescope}  using the Wide Field Planetary Camera 2 (WFPC2) in 2007 as part of  the SNAPshot programme GO-11103 (PI: Ebeling). This observation established  MACSJ0416  as a powerful gravitational lens which led to its inclusion in the CLASH programme \citep[PI: Postman;][]{postman12}. Hence, MACSJ0416 was observed with HST in 2012 for a total  of 20 orbits across 16 passbands, from the UV to the near-IR. Table~\ref{HFFdata} lists details of these ACS and WFC3 observations, which were used for the  pre-HFF analysis of MACSJ0416.  All mass models based on pre-HFF data \citep[][]{johnson14,coe14,richard14} are publicly available\footnote{http://archive.stsci.edu/prepds/frontier/lensmodels/}.

\subsection{Hubble Frontier Fields Data}
MACSJ0416 is the second cluster to be observed in the HFF program (GO/DD 13496). Observations with ACS were performed from January 5$^{th}$ to February 9$^{th}$ 2014 in three filters (F435W, F606W, and F814W) for  a total exposure time of 20, 12, and 48 orbits, respectively. A summary of these observations is provided at the end of Table~\ref{HFFdata}.  At the time of this writing,  imaging of the cluster with WFC3 for the HFF programme had yet to commence.  

We applied basic data-reduction procedures to the HFF/ACS data, using {\tt HSTCAL} and the most recent calibration files. Individual frames were co-added using {\tt Astrodrizzle} after registration to a common ACS reference image using {\tt Tweakreg}. After an iterative process, we achieve an alignment accuracy of 0.1 pixel. Our final stacked images have a pixel size of 0.03\arcsec.

\subsection{Chandra X-ray Data}
MACSJ0416 was observed with the Advanced CCD Imaging Spectrometer (ACIS-I) on board the \emph{Chandra} X-ray Observatory on 2009-06-07 for 16 ks (ObsID 10446, PI: Ebeling), and on 2014-06-09 for 37 ks (ObsID 16237, PI: Jones). We process these archival data using CIAO v4.6 and CALDB v4.5.9, merging them into a single 53 ks observation. After examining the light curve of the accumulated count rate for periods of enhanced particle background, we  extract a raw image of the cluster in the 0.7--7 keV band and use the CIAO tool \texttt{mkexpmap} to compute an effective exposure map, taking vignetting effects into account.  The raw image, which preserves the recorded photon statistics, is adaptively smoothed using \texttt{asmooth} \citep{ebeling06}, requiring 3$\sigma$ significance of all features with respect to the local background. 
\label{chandra_obs}

\subsection{Spectroscopic and Photometric Redshifts}
\label{zspec_zphot_obs}
More than 100 spectroscopic galaxy redshifts are available within the field of MACSJ0416.  Spectroscopic redshifts
from \cite{ebeling14} are complemented by redshifts obtained for VLT programme 186.A--0798 (Balestra et al., in preparation).  We also make use of the catalogue of photometric redshifts derived by the CLASH team from HST imaging in 16 passbands (second block of entries in Table~\ref{HFFdata}) utilising the Bayesian Photometric Redshift (BPZ) programme \citep{benitez04,coe06}.  We use of all of these redshifts to select background galaxies for our weak-lensing catalogue, as well as for the identification (and removal) of cluster members (see Sect.~\ref{WLcatalogue} for details).

\section{Strong-Lensing Analysis: Revisiting Multiple Images}
\label{SLanalysis}
Since our HFF strong-lensing analysis of MACSJ0416 has already been presented in \cite[][hereafter J14]{jauzac14}, we here provide only a brief synopsis of the mass model derived and the main results.

Before the HFF observations of MACSJ0416, \cite{zitrin13a} identified 23 multiple-image systems (corresponding to a total of 70 individual images) in the CLASH data \citep[][]{postman12}. The identifications of 10 of these (comprising 36 individual images) were considered less robust. In \cite[][ hereafter R14]{richard14}, we  included only the most robust systems as well as a few candidate systems showing clear counter-images at locations predicted by our preliminary strong-lensing analysis. Our final list contained 17 multiply imaged galaxies (47 individual images). Thanks to the unprecedented depth of the HFF data, J14 was able to dramatically improve these numbers, by discovering 51 new multiple-image systems, bringing our final list of identifications to 68 multiply imaged galaxies, with a total of 194 individual images. Spectroscopic confirmation, however, has so far been obtained for only 9 systems. The full list of these systems is given in J14.

Using a subset of the 57 most securely identified multiple-image systems, we built a strong-lensing parametric mass model using the publicly available \textsc{Lenstool}\footnote{http://projects.lam.fr/repos/lenstool/wiki} software. The resulting best-fit model comprises two cluster-scale dark-matter halos and 98 galaxy-scale halos. The parameters describing this best-fit mass model are listed in Table~\ref{table_SLparam}. We also provide a mass estimate for both components within 20$\arcsec$ ($\sim$100~kpc).

This model predicts image positions to within an $RMS$ error of 0.68\arcsec, an improvement in precision of almost a factor of two over pre-HFF models of this cluster. The total mass enclosed in the multiple-image region is $M_{SL}{\rm (R<320~kpc)} = (3.26 \pm 0.03)\times 10^{14}$M$_{\odot}$. This measurement offers a three-fold improvement in precision and drives the statistical mass uncertainty below 1\% for the first time in any cluster. Finally, the  statistical uncertainty in the median magnification has been lowered to 4\%. The resulting high-precision magnification map of this powerful cluster lens immediately improves the constraints on the luminosity function of high-redshift galaxies lensed by this system. 

For more details on our HFF strong-lensing analysis of MACSJ0416, we refer the reader to J14, where the methodology and mass measurements are described in details. 

\begin{table}
\begin{center}
\begin{tabular}[h!]{cccc}
\hline
\hline
\noalign{\smallskip}
Component  & C1 & C2 & L$^*$ elliptical galaxy \\
\hline
$\Delta$ \textsc{ra}  & -4.5$^{+0.7}_{-0.6}$  &  24.5$^{+0.5}_{-0.4}$ & --  \\
$\Delta$ \textsc{dec} & 1.5 $^{+0.5}_{-0.6}$  & -44.5$^{+0.6}_{-0.8}$ &  --  \\
$e$ & 0.7 $\pm$0.02  & 0.7$\pm$0.02 & -- \\
$\theta$ & 58.0$^{+0.7}_{-1.2}$  & 37.4$\pm$0.4 & -- \\
r$_{\mathrm{core}}$ (\footnotesize{kpc}) & 77.8$^{+4.1}_{-4.6}$  & 103.3$\pm$4.7  & [0.15] \\
r$_{\mathrm{cut}}$ (\footnotesize{kpc}) & [1000] & [1000] & 29.5$^{+7.4}_{-4.3}$ \\
$\sigma$ (\footnotesize{km\,s$^{-1}$}) &  779$^{+22}_{-20}$ & 955$^{+17}_{-22}$ & 147.9$\pm$ 6.2 \\
M (\footnotesize{$10^{13}$ M$_{\odot}$}) & 6.02$\pm$0.09 & 6.12$\pm$0.09 \\
\noalign{\smallskip}
\hline
\hline
\end{tabular}
\caption{Best-fit PIEMD parameters for the two large-scale dark-matter halos. 
Coordinates are quoted in arcseconds with respect to $\alpha{=}64.0381013$ deg, $\delta{=}-24.0674860$ deg (yellow cross in Fig.~\ref{m0416cons}).
Errors correspond to the $1\sigma$ confidence level. Parameters in brackets are not optimised.
The reference magnitude for scaling relations is $m_{\rm F814W} = 19.8$. Masses are quoted within an aperture of 20$\arcsec$ ($\sim$100~kpc).
}
\label{table_SLparam}
\end{center}
\end{table}

\section{Weak-Lensing Constraints}
\label{WLcatalogue}

In this Section, we summarise our analysis methodology and then discuss in particular the enhancements to our technique wrought by significant improvements in the data quality brought about by the HFF observations. A more detailed description of the method used to generate the weak-lensing background-galaxy catalogue is presented in \cite{jauzac12} (hereafter J12). 

\subsection{The ACS Source Catalogue}
\label{acscat}
\begin{figure}
\hspace*{-3mm}\includegraphics[width=0.5\textwidth]{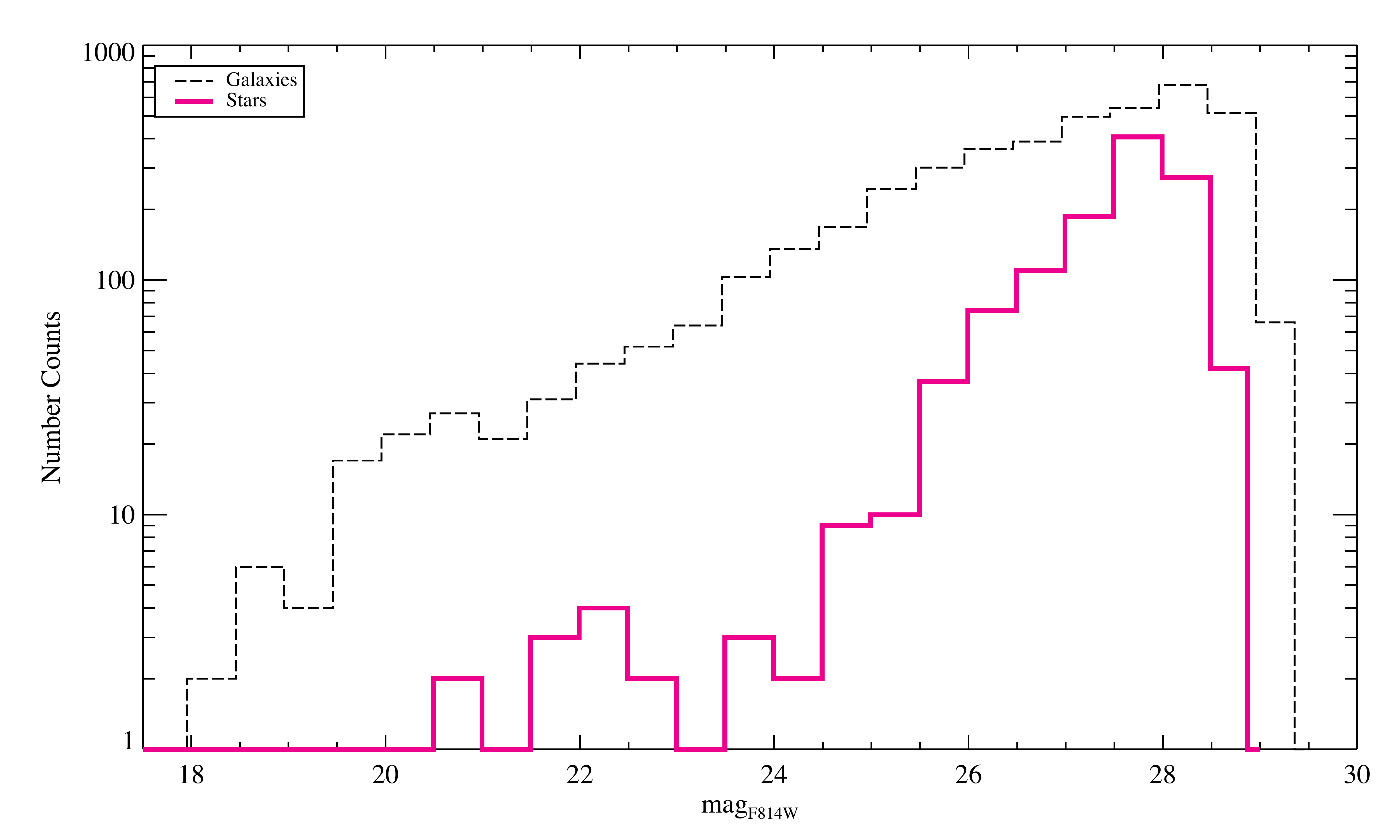}
\caption{Magnitude distribution of the sources identified as galaxies (dashed black) and stars (magenta) resulting from our \textsc{SEXTRACTOR} detection.}
\label{histo_mag_galstar}
\end{figure}

Our weak-lensing analysis is based on shape measurements in the ACS/F814W band. Following a method developed for the analysis of data obtained for the COSMOS survey, and described in \cite{leauthaud07} (hereafter L07), the SE\textsc{xtractor} photometry package \citep{BA96} is used to detect sources with the \textit{`Hot-Cold'} method \citep[][L07]{rix04}. It consists of running {\tt Sextractor} twice: first with a configuration optimised to detect only the brightest objects (the \textit{cold} step), and then a second time with a configuration optimised to detect the faint objects that contain most of the lensing signal (the \textit{hot} step). The resulting catalogue is then cleaned by removing spurious or duplicate detections using a semi-automatic algorithm that defines polygonal masks around stars or saturated pixels. Galaxies are distinguished from stars by examining the distribution of objects in the magnitude (MAG$\_$AUTO) vs peak surface brightness (MU$\_$MAX) plane (see L07 $\&$ J12 for more details). Finally, the drizzling process introduces pattern-dependent correlations between neighbouring pixels, which artificially reduces the noise level of co-added drizzled images. Care must be taken to correct for this effect. Because we have used the same Drizzle pixelfrac and convolution kernel parameters as L07, we apply the same remedy as L07 by simply scaling up the noise level in each pixel by the same constant F$_{A} \approx$ 0.316, defined by \cite{casertano00}. The resulting catalogue comprises 4296 sources identified as galaxies and 1171 sources identified as point sources (stars) within a magnitude limit of $m_{\rm F814W} = 29.5$. Fig.~\ref{histo_mag_galstar} shows the magnitude distribution of the detected galaxies and stars.

\begin{figure}
\hspace*{-3mm}\includegraphics[width=0.5\textwidth]{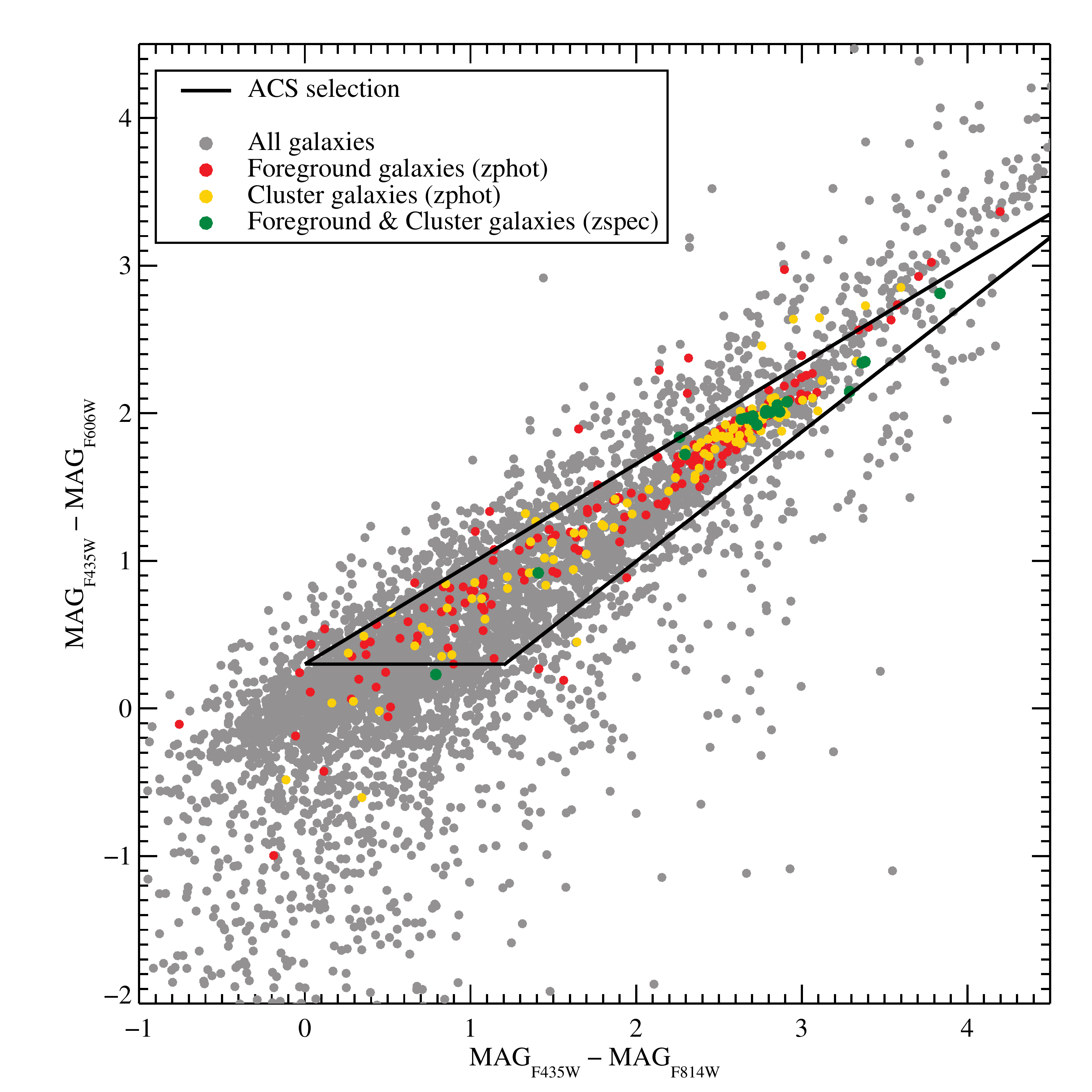}
\caption{Colour-colour diagram ($m_{\rm F435W}-m_{\rm F814W}$) vs ($m_{\rm F435W} - m_{\rm F606W}$) for objects within the HFF/ACS image of MACSJ0416. Grey dots represent all galaxies in the study area. Unlensed galaxies diluting the shear signal are marked by different colours: galaxies spectroscopically confirmed as cluster members or foreground galaxies (green); galaxies classified as foreground objects because of their photometric redshifts (red); and galaxies classified as cluster members via photometric redshifts (yellow). The solid black lines delineate the colour-cut defined for this work to mitigate shear dilution by unlensed galaxies.}
\label{cc_diagram}
\end{figure}

\begin{figure}
\hspace*{-3mm}\includegraphics[width=0.5\textwidth]{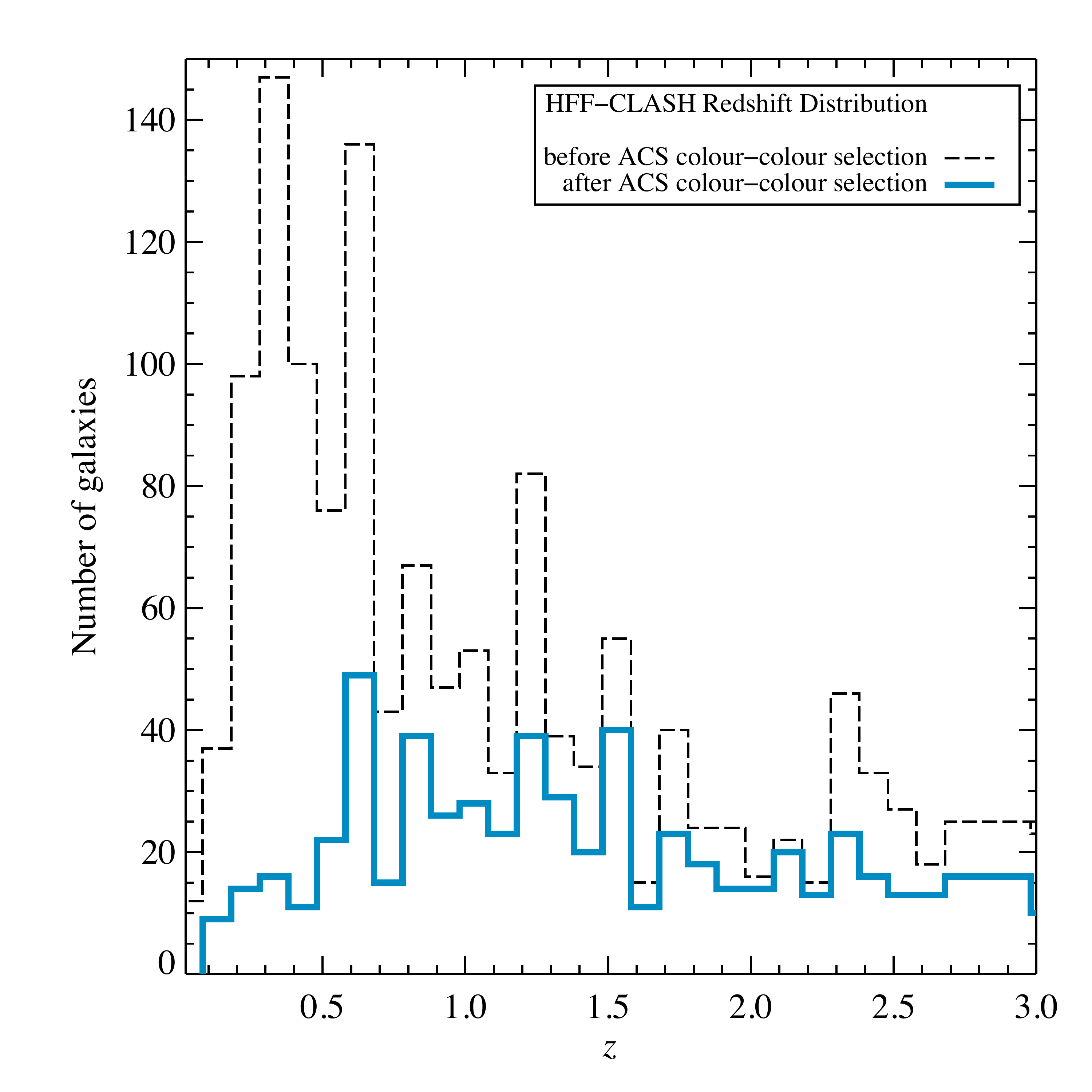}
\caption{Redshift distribution of all galaxies with $m_{\rm F435W}$, $m_{\rm F606W}$, and $m_{\rm F814W}$ photometry from HFF observations, that have photometric or spectroscopic redshifts (dashed black histogram). The cyan histogram shows the redshift distribution of galaxies classified as background objects using the colour-colour criterion illustrated in Fig.~\ref{cc_diagram}.}
\label{Nz_ccselect}
\end{figure}

Since only galaxies behind the cluster are gravitationally lensed, the presence of cluster members and foreground galaxies dilutes the observed shear and reduces the significance of all quantities derived from it.  Identifying and eliminating as many of the contaminating unlensed galaxies is thus crucial.  As a first step, we identify cluster galaxies with the help of the catalogue of photometric redshifts compiled by the CLASH collaboration, and the spectroscopic redshifts mentioned in Sect.~\ref{zspec_zphot_obs}. All galaxies with photometric redshift 0.35$<$z$_{phot}$$<$0.44 are considered to be cluster galaxies. The spectroscopic cluster membership criterion is defined by
$$z_{\rm cluster} - dz < z < z_{\rm cluster}+ dz ,$$
where $z$ is the spectroscopic redshift of the considered galaxy, $z_{\rm cluster} = 0.3979$ is the systemic redshift of the cluster, and $dz = 0.0104$ is the 3$\sigma$ cut defined by the colour-magnitude selections presented in Sect.~\ref{CM_model}. Only 30\% of the sources in our ACS object catalogue have a photometric redshift. Of these 30\%, 17\% are identified as cluster members or foreground sources following the aforementioned selection criteria. Due to the large difference in depth between the CLASH and HFF observations, the photometric redshift catalogue is not sufficient to identify all the unlensed contaminants in our catalogue. Therefore, taking advantage of the 3 HFF ACS-band imaging, we use a colour-colour diagram to identify foreground and cluster members (Fig.~\ref{cc_diagram}). Using galaxies with photometric or spectroscopic redshifts, we identify the region in colour-colour space that is dominated by unlensed galaxies (foreground galaxies and cluster members) and define its boundaries by $m_{\rm F435W}-m_{\rm F814W}$ $<$ 0.67776 ($m_{\rm F435W}-m_{\rm F606W}$) + 0.3; $m_{\rm F435W}-m_{\rm F814W}$ $>$ 0.87776 ($m_{\rm F435W}-m_{\rm F606W}$) - 0.76; $m_{\rm F435W}-m_{\rm F814W}$ $>$ 0.3. We consequently remove all objects within this region from our analysis.
Fig.~\ref{Nz_ccselect} shows the galaxy redshift distribution before and after this F435W-F606W-F814W colour-colour selection. This selection is very efficient at removing cluster members and foreground galaxies at $z\leqslant$0.44 --- for the subset of our galaxies that have redshifts, 88\% of the unlensed population are eliminated.

\begin{figure}
\hspace*{-3mm}\includegraphics[width=0.5\textwidth]{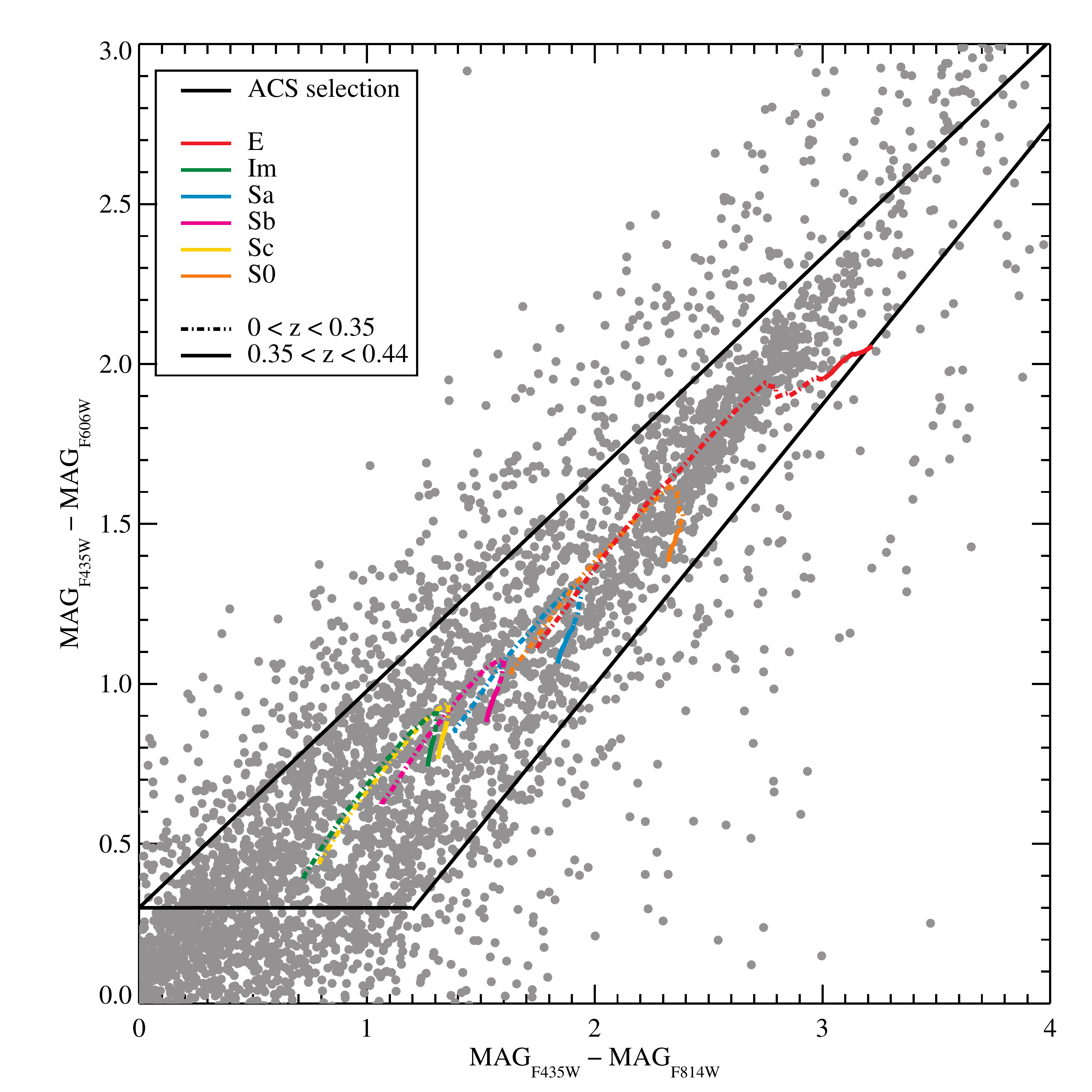}
\caption{Colour-colour diagram ($mag_{F435W}-mag_{F814W}$) vs ($mag_{F435W} - mag_{F606W}$) as in Fig.~\ref{cc_diagram}. 
The solid black lines delineate the colour-cut defined for this work. The different spectral templates predicted from \citet{CWW} and \citet{kinney96} and the theoretical ones from \citet{BC03} are marked by different colours: elliptical galaxies (red); magelanic irregulars (green); spiral Sa galaxies (cyan); spiral Sb (magenta); spiral Sc (yellow); and S0 galaxies (orange).
The dash--dot curves correspond to a redshift range $0<z<0.35$ (foreground galaxies), and the solid ones to $0.35<z<0.44$ (cluster members).}
\label{cctracks}
\end{figure}

The final validation of our colour-colour selection is done by predicting the colours expected from spectral templates at the redshift of the cluster or in the foreground. We use empirical templates from \citet{CWW} and \citet{kinney96} as well as theoretical templates from \cite{BC03} for various galaxy types in the Hubble sequences (ranging from Elliptical to SB) and starburst galaxies. The location of the colour-colour tracks at $z<0.44$ agree well with our selection region as shown in Fig.~\ref{cctracks} for the Bruzal \& Charlot model.

\subsection{Shape Measurements \& Lensing Cuts}
Measurements of galaxy shapes are performed using the RRG method \citep{rhodes00}. This method has been developed for the analysis of data obtained from space, featuring a small, diffraction-limited point-spread function (PSF). It decreases the noise in the shear estimators by correcting each moment of the PSF linearly and only dividing them at the very end to compute an ellipticity. \cite{rhodes07} demonstrated that the ACS PSF  is not as stable as one might expect from a space-based camera. Its size and ellipticity vary considerably on time-scales of weeks due to telescope `breathing', which induces a deviation from the nominal focus and thus from the nominal PSF which becomes larger and more elliptical. To overcome this problem, \cite{rhodes07} created a grid of simulated PSF images. From a comparison of the ellipticity of $\sim$20 stars in each image to the ellipticities of these model images, the effective focus of the observation can be determined. PSF parameters are then interpolated using the method presented in \cite{massey02}. This technique was used in L07 and J12. However, it has since been shown that PSF variations occur even between subsequent exposures, and thus a modeling of the PSF for each epoch results in a more accurate estimation of the correction to apply to shear estimations (Harvey et al., \textit{in prep.}).

In order to handle the multi-epoch images of MACSJ0416 we adapted the RRG pipeline (L07) to model the average PSF at the position of each galaxy in the stacked image (see Harvey at al \textit{in prep.} for more details). To this end, we first locate the positions of the stars in the reference frame of the drizzled image using {\tt Sextractor} and  both the magnitude -- size and magnitude -- MU\_MAX diagrams. We measure the second- and fourth-order moments of these stars from {\em each} exposure and compare them to the Tiny Tim model for the F814W band. Using the best-fitting TinyTim PSF model, we then interpolate the PSF to the galaxy positions, rotate the moments such that they are in the reference frame of the stacked image, and then take an average over the stack. (Note that our PSF model thus depends on the number of exposures covering a given area and is not necessarily a continuous function across the field.)  Since we know the number of exposures that contribute to the image of each galaxy, we can discard shear estimates of galaxies that have fewer than 3 exposures. 
Doing so removes all galaxies near the edge of the field and along chip boundaries. 

The RRG method returns three parameters: $d$, a measure of the galaxy size, as well as $e_1$ and $e_2$, the two components of the ellipticity vector $e = (e_1, e_2)$, defined as 
$$d = \sqrt{\frac{1}{2} (a^{2} + b^{2})}$$
$$e = \frac{a^{2} - b^{2}}{a^{2} + b^{2}}$$
$$e_{1} = e cos(2\phi)$$
$$e_{2} = e sin(2\phi) .$$
Here $a$ and $b$ are the major and minor axes of the background galaxy, and $\phi$ is the orientation angle of the major axis. The ellipticity $e$ is then calibrated by a factor called shear polarisability, $G$, to obtain the shear estimator $\tilde{\gamma}$:
\begin{equation}
\tilde{\gamma} = C \frac{e}{G}.
\label{eqn:shear}
\end{equation}
We use the same global measurement of the shear susceptibility $G$ as in L07:
$$G = 1.125 + 0.04 \arctan\frac{S/N - 17}{4} .$$
Finally, $C$ in Eq.\ref{eqn:shear} is the calibration factor, determined using a set of simulated images similar to those used by STEP \citep{heymans06, massey07} for COSMOS images, and is given by $C = (0.86^{+0.07}_{-0.05})^{-1}$ (see L07 for more details).

The last step in constructing the weak-lensing catalogue is to exclude galaxies whose shape parameters are so ill-determined that including them would increase the noise in the shear measurements more than the shear signal. These cuts are the same as the ones used in L07 and J12, and are quoted here for clarity:
\begin{itemize}
\item
Threshold in the estimated detection significance:
$$\frac{S}{N} = \frac{FLUX\_AUTO}{FLUXERR\_AUTO} > 4.5 ;$$
\item
Threshold in the total ellipticity:
$$e = \sqrt{e_{1}^{2} + e_{2}^{2}} < 1 ;$$
\item
Threshold in the size, as defined by the RRG $d$ parameter :
$$3.6 < d < 30~\textrm{pixels}.$$
\end{itemize}

As explained in J12, the requirement that the galaxy ellipticity be less than unity may appear trivial and superfluous. In practice it is meaningful though since the RRG method allows measured ellipticity values to be greater than 1 due to noise, although ellipticity is by definition restricted to $e \leqslant 1$. The lower limit in the RRG size parameter $d$ aims to eliminate sources with uncertain shapes, since PSF corrections and thus credible shape measurements become increasingly difficult as the size of a galaxy approaches that of the PSF. The upper limit in $d$ aims to eliminate sources with a size similar to large elliptical cluster members. In addition to applying the aforementioned cuts, and in order to ensure an unbiased mass reconstruction while combining strong and weak lensing, we also remove all background galaxies located in the multiple-image (strong lensing) region, which can be approximated by an ellipse aligned with the cluster elongation as predicted by the strong-lensing model ($a=75\arcsec$, $b=36\arcsec$, $\theta = 135\deg$, $\alpha=64.0351$ deg, $\delta=-24.0745$ deg). 

\begin{figure}
\hspace*{-3mm}\includegraphics[width=0.5\textwidth]{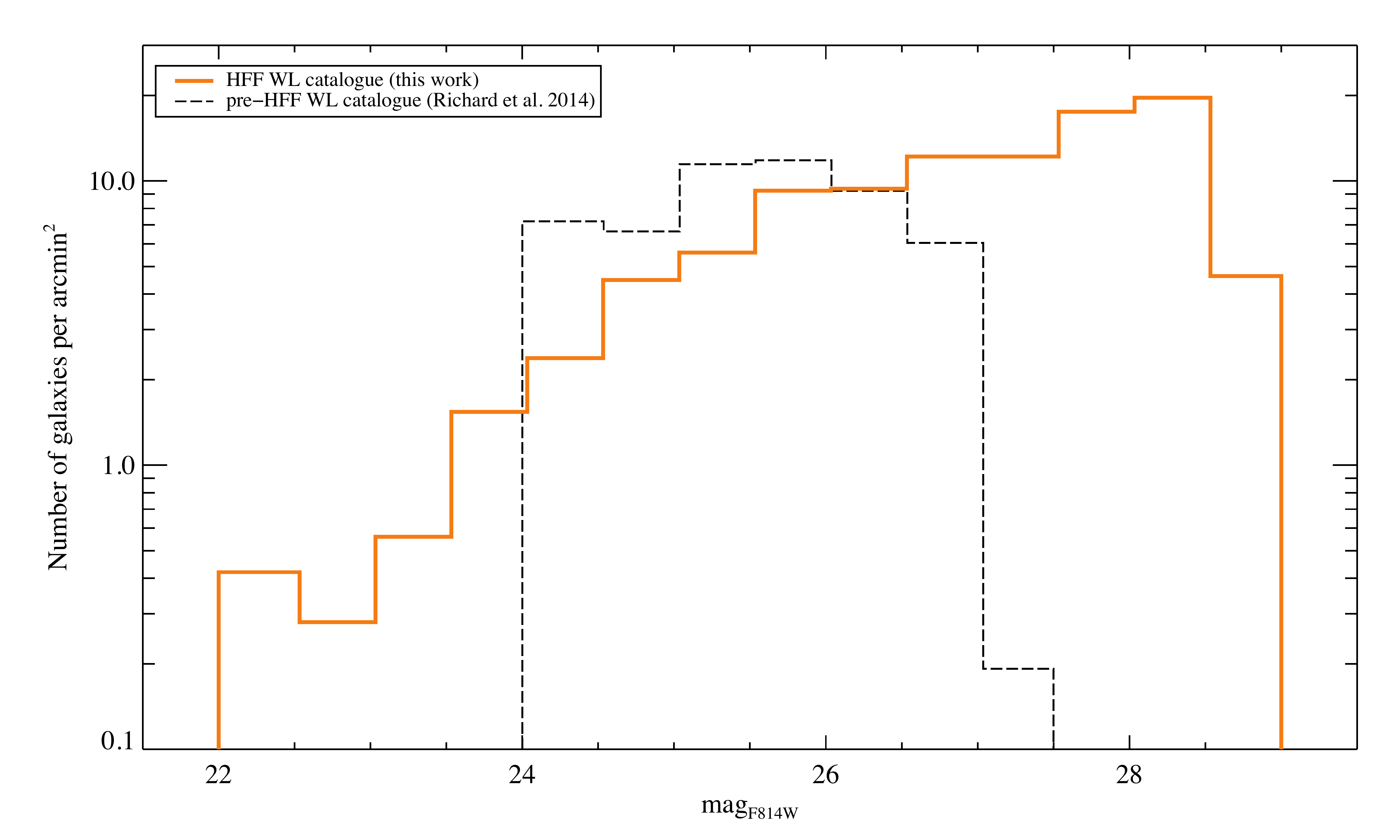}
\caption{Magnitude distribution of background galaxies selected for our pre-HFF analysis per arcmin$^{2}$ (dashed black line, R14), and for this work (orange). }
\label{comp_WLcats}
\end{figure}

Fig.~\ref{comp_WLcats} compares the magnitude distribution of selected background galaxies for the pre-HFF data (R14) and the HFF data. The HFF-based catalogue extends to ACS-F814W magnitudes of 29, two magnitudes fainter than the pre-HFF dataset. Note that the shown distributions differ also at lower magnitudes. Owing to the greatly increased depth of the HFF data compared to those obtained for the CLASH programme, the contamination by faint foreground and cluster galaxies is much increased too. As a consequence, our colour-colour selection is more drastic and removes more objects in the magnitude range $24<m_{\rm F814W}<26$. 13 galaxies with $m_{\rm F814W}<24$ are still included in our HFF catalogue that were removed from the pre-HFF catalogue of R14. The reason is the different colour-colour selection employed by R14, which was less efficient and required a magnitude cut at $m_{\rm F814W}=24$ to remove bright objects.
The depth of the HFF images also causes more stars to be saturated, requiring the size of the corresponding  masks to be increased; in total $\sim$40\% of the ACS surface is masked out as a result. Our final weak-lensing catalogue is composed of 714 background galaxies, corresponding to a density of $\sim$ 100~galaxies.arcmin$^{-2}$. Compared to the catalogue generated by our pre-HFF analysis (R14), the density of weakly lensed galaxies has almost doubled.

\section{Mass Modeling: Combining Strong- and Weak-Lensing Constraints}
\label{gridmethod}
As, in this study, we aim to detect dark-matter-dominated structures outside the strong-lensing region, we add a grid-based model to the parametric lens model described in Sect.~\ref{SLanalysis}.

\subsection{Grid Method: Combining Strong \& Weak Lensing Constraints}
\label{comb_slwlmodel}
In previous work, we modelled the matter distribution with a set of Radial Basis Functions (RBFs) located at the nodes of a multiscale grid, which covered an area slightly larger than observed \citep[][]{jullo09,jauzac12,jullo14}. 
We complemented this grid-based model with dPIE potentials \citep{eliasdottir07} to account for the  lensing contribution of  cluster members. 

In this work, we adopt a slightly different approach. We keep the parametric model described in section~\ref{SLanalysis} fixed at the best-fit values and then estimate the RBF amplitudes from the WL constraints. By doing so, we appropriately weigh the SL constraints and do not account for them twice. Indeed, an attempt at optimizing the RBF amplitudes with both strong- and weak-lensing constraints failed to produce physically meaningful results as the strong-lensing signal completely dominated the optimization process and essentially overwhelmed the weak-lensing data. The parametric model contains two cluster-scale halos and 146 galaxy-scale halos, as we extend our analysis to the full ACS field of view (see next subsection).  We add a uniform grid of RBFs to these main mass components. Each RBF is axi-symetric, fixed in position and size, and only its amplitude varies, as if they were pixels in an image. The radial profile of each RBF is modelled with a dPIE potential \citep{eliasdottir09}. The core radius $s$ is set to the distance between an RBF and its closest neighbour, and the cut radius $t$ is assumed to be three times the core radius \cite{jullo09}.

We tried different prescriptions for the grid resolution. The optimum solution was achieved using a uniform grid with 2741 RBFs separated from each other by $s=5.5\arcsec$ (see Sect.~\ref{gridres} for more details).  In addition, we found it necessary to remove the RBFs in the central strong-lensing area since, due to a lack of weak-lensing constraints, the reconstruction in this region was very noisy. The reconstruction is smooth, thanks to overlapping RBFs.

We sum the contribution of the two components of our model to the observed ellipticity as follows:

\begin{equation}
\label{eq:shearmat}
\bold{e_m} = M_{\gamma v} \bold{v} + \bold{e_{\rm param}} + \bold{n}\;,
\end{equation}

\noindent Here the vector $\bold{v}$ contains the amplitudes of the 2741 RBFs,  vector $\bold{e_m} = [\bold{e_1}, \bold{e_2}]$ contains the individual shape measurements of the weak-lensing sources,  $\bold{e_{\rm param}}$ is the fixed ellipticity contribution from the parametric model. The vector $\bold{n}$ represents the Gaussian noise in the shape measurements, i.e., the intrinsic ellipticity of galaxies and the noise in our measurements of their shapes.  The transformation matrix $M_{\gamma v}$ contains the cross-contribution of each individual RBF to each individual weak-lensing source. Each shear component in scaled by the ratio of the distances between each individual source $S$, the cluster $L$, and the observer $O$. The elements of the $M_{\gamma v}$ matrix for the two shear components are

\begin{eqnarray}
\label{eq:dshear1}
\Delta_{1}^{(j,i)} &= &\frac{D_{LSi}}{D_{OSi}}\ \Gamma_{1}^i(|| \theta_i - \theta_j ||,\ s_i,\ t_i) , \\
\Delta_{2}^{(j,i)} &= &\frac{D_{LSi}}{D_{OSi}}\ \Gamma_{2}^i(|| \theta_i - \theta_j ||,\ s_i,\ t_i) . 
\end{eqnarray}

\noindent  where analytical expressions for $\Gamma_1$ and $\Gamma_2$ are given in \citet[][Equation\ A8]{eliasdottir09}. Note that the shear in the cluster can be large, and the assumption shown in Eq.~\ref{eq:dshear1} may not be strictly valid. However, since the parametric model accounts for most of the lensing effect, the contribution to the grid-based model originates primarily in the weak-lensing regime.

\subsection{Modeling of Cluster Members}
\label{CM_model}
Complementing our grid of RBFs, we add the contributions from 146 cluster member galaxies (presented in Sect.~\ref{comb_slwlmodel}), modelled again as dPIE potentials and selected following the method presented in R14. 
We define cluster members to be those galaxies that fall within 3$\sigma$ of a linear model of the cluster red sequence in both the ($m_{\rm F606W} - m_{\rm F814W}$) vs $m_{\rm F814W}$ and the ($m_{\rm F435W} - m_{\rm F606W}$) vs $m_{\rm F814W}$ colour-magnitude diagrams. The magnitude limit of this sample is $m_{\rm F814W} = 23.4$. 

These galaxies are then inserted in the model as small-scale pertubators. Their cut radius and velocity dispersions are fixed, and scaled from their luminosities in HST/ACS F814W-band. We derive L$^*$ in our filter of observation based on the K$^*$ magnitudes obtained by \cite{lin06} as a function of cluster redshift. Cut radius and velocity dispersion are then scaled relative to an $m^* = 19.76$ galaxy with velocity dispersion $\sigma^* = (119\pm20)$ km s$^{-1}$ and cut radius $r_{cut}^*  = (85\pm20)$ kpc.

\subsection{Priors and MCMC sampling}
We sample the huge parameter space with the MassInf algorithm implemented in the Bayesys library \citep{skilling98}, itself implemented in Lenstool, and described in \citet{jullo14}.  In summary, this algorithm uses the Gibbs sampling approach in which, at each iteration, the most significant RBFs are first identified and then adjusted in amplitude to fit the ellipticity measurements. The number of significant RBFs is a prior of MassInf, although we have shown in \cite{jullo14} that it has little impact on the reconstruction. We set the initial number of significant RBFs to 2\%, and the algorithm converged to about 4\%.
In contrast to previous analyses, we here do not assume that the resulting mass distribution has to be positive everywhere. We found that incorporating such a prior introduces a spurious bias favouring positive values of the mass-sheet degeneracy.

The objective function is a standard likelihood function, in which noise is assumed to be Gaussian. The algorithm returns a large number of MCMC samples, from which we can estimate mean values and errors on different quantities (mass density field, amplification field, etc).

\begin{figure*}
\includegraphics[width=1.0\textwidth,angle=0.0]{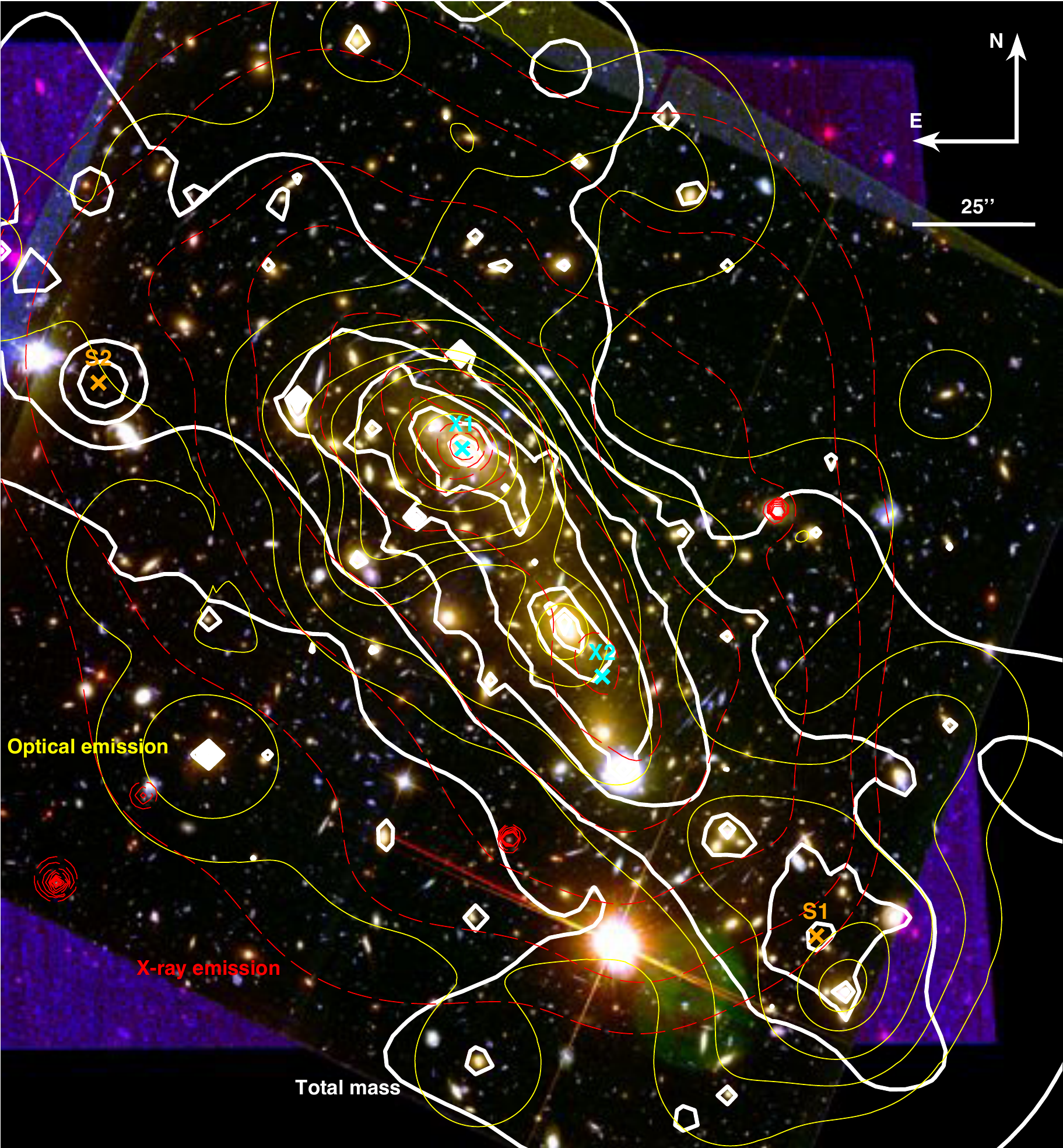}
\caption{Composite colour image of the galaxy cluster MACSJ0416 created from HST/ACS images in the F814W, F606W, and F435W passbands. Mass contours from our gravitational-lensing analysis are shown in bold white, contours of the adaptively smoothed X-ray surface brightness in the 0.5--7 keV band as observed with \emph{Chandra} are shown in dashed red, while the light distribution is delineated by yellow contours. Substructures S1 and S2 are marked with orange crosses while the two X-ray peaks, labeled X1 and X2, are  marked by cyan crosses.}
\label{m0416cons}
\end{figure*}

\subsection{Redshift Estimation for Background Sources}
Of the 714 background galaxies in our catalogue, 236 have a photometric redshift estimated from the CLASH data that allow us to isolate background galaxies. We found the following function to provide a good description of the redshift distribution of these background galaxies:
\begin{equation}
    \mathcal{N}(z) \propto e^{- (z /z_0)^\beta}
\end{equation}
\noindent with $\beta = 1.84$ and a median redshift $<z>$ $= 1.586 = 0.56\, z_0$ \citep[][]{gilmore09,natarajan97}. 

In addition, we split the catalog into a bright and a faint subsample at the median magnitude $m_{\rm F814W} = 26.4$. 
Within the uncertainties given by the number statistics, the resulting two histograms have the same slope.
Since \textsc{Lenstool} allows each source to have its own redshift, we randomly draw (during the initialization phase) redshifts from the fitted redshift distribution for all  galaxies without spectroscopic or photometric redshift.

\section{Results}
We now present our results for the properties of both dark and luminous matter in MACSJ0416, beginning with the distribution of the total gravitating mass as reconstructed by our lensing analysis.
All masses for the cluster, as well as density profiles, are measured with respect to the position of the brightest cluster galaxy (BCG), i.e., $\alpha=$04:16:09.144, $\delta=$-24:04:02.95.
\subsection{Distribution of Total Mass}
\label{totalmass}
\begin{figure}
\hspace*{-3mm}\includegraphics[width=0.5\textwidth]{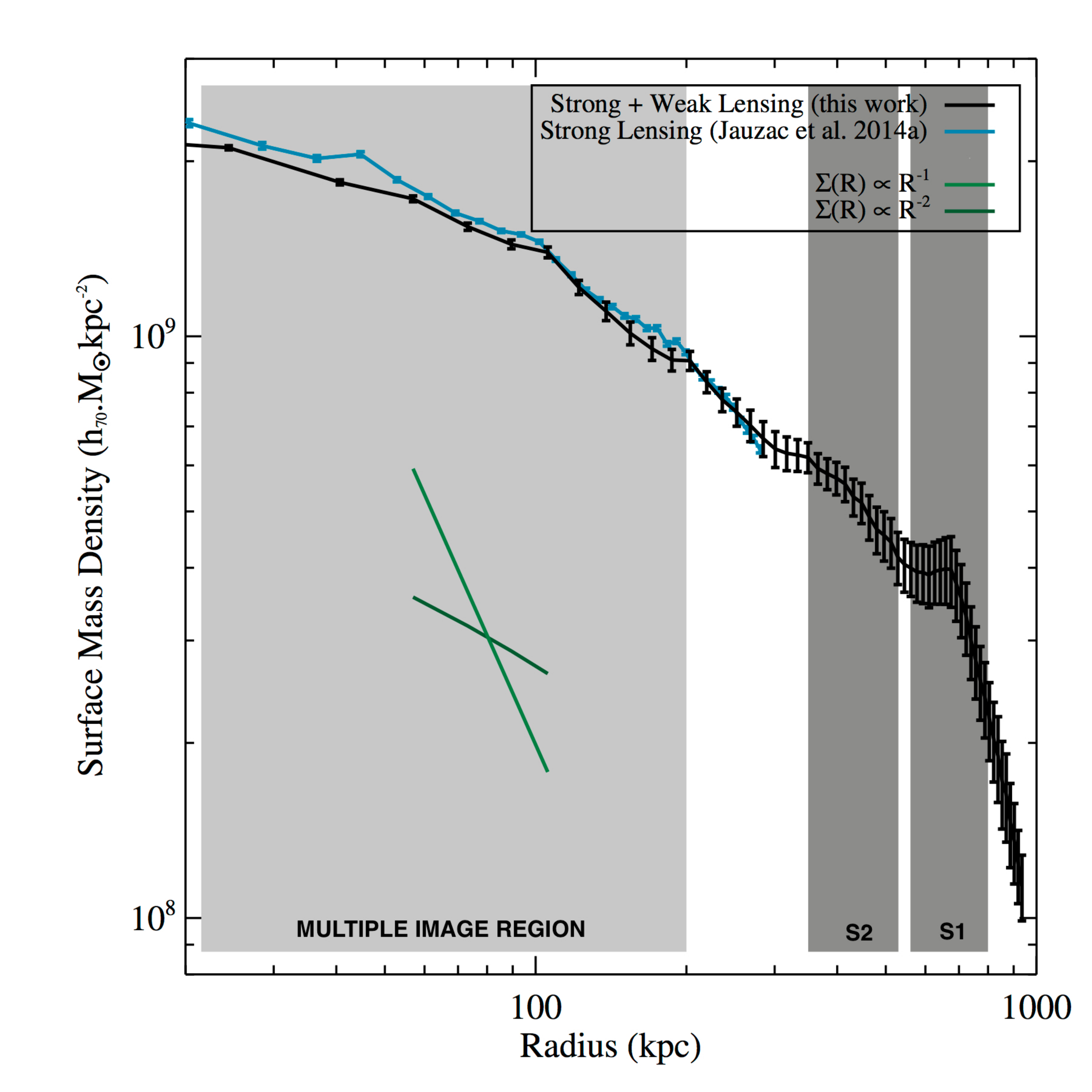}
\caption{Surface density profile obtained with our comprehensive gravitational-lensing analysis (black). For comparison, we show the surface density profile obtained by J14 based exclusively on strong-lensing features (cyan). The shaded light grey area marks the region within which multiple-image systems were found. The two dark grey shaded areas mark the region of substructures S1 and S2. The two green lines represent two different slopes: $R^{-1}$ and $R^{-2}$.}
\label{densprof}
\end{figure}

\subsubsection{Grid Resolution}
\label{gridres}
To assess whether the grid of RBFs presented in Sect.~\ref{gridmethod} is optimally suited to describe MACSJ0416, we compared the Bayesian evidence resulting from the optimisation for grids of different resolution. We remind the reader that the logarithmic Bayesian evidence is given by
$$ \log(E) = \int^{1}_{0} <\log(L)>^{\lambda} d\lambda$$,
where the average is computed over a set of 10 MCMC realizations at any given iteration step $\lambda$, and the integration is performed over all iterations $\lambda_{i}$ from the initial model ($\lambda=0$) to the best-fit result ($\lambda=1$). The increment $d\lambda$ depends on the variance between the 10 likelihoods computed at a given iteration, and on a convergence rate that we set equal to 0.1 (see Bayesys manual for details). The increment gets larger as the algorithm converges towards 1.

To test the impact of higher resolution, we created a uniform grid containing 3883 RBFs separated by $s=3.3\arcsec$. We found the resulting model added noise where no structures are detected in the optical. The Bayesian evidence obtained for this high-resolution grid is $\log(E) = -267$, compared to $\log(E) = -251$ for the grid described in Sect.~\ref{gridmethod}. We also explored a low-resolution grid, containing 722 RBFs separated by $s=11.1\arcsec$. The optimization using this grid resulted in a  better Bayesian evidence of $\log(E) = -229$, and a similar $\chi^2$. However, all structures were poorly resolved. Since, in addition, the signal-to-noise ratios in the sub-structures were equivalent, we discarded this low-resolution model.

\subsubsection{Comparison with Previous Analyses}

\begin{figure}
\hspace*{-3mm}\includegraphics[width=0.5\textwidth]{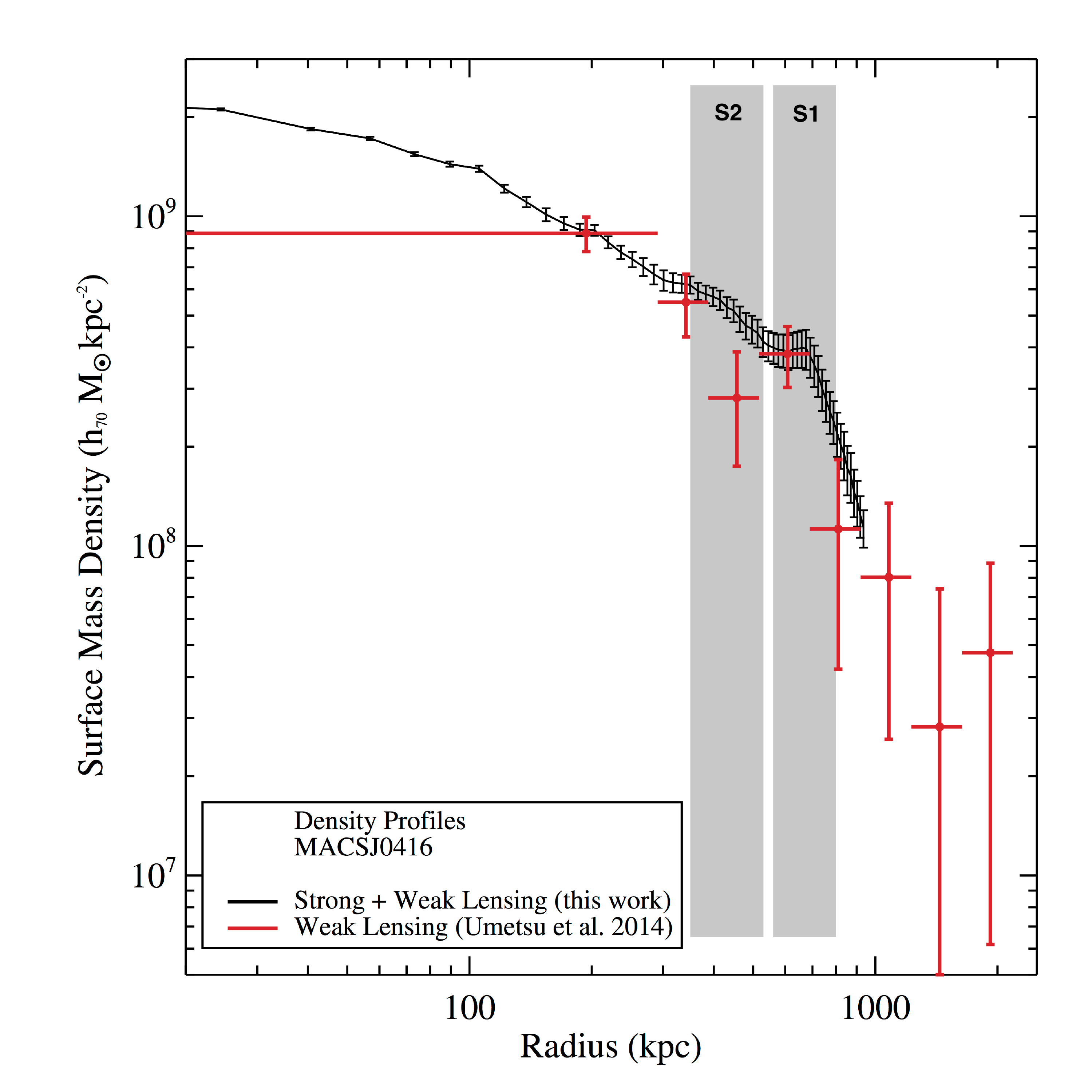}
\caption{Surface density profile obtained with our comprehensive gravitational-lensing analysis (black). For comparison, we show the surface density profile obtained by \citet{umetsu14} based exclusively on weak-lensing features (red) derived thanks to Subaru observations (Keiichi Umetsu private communication). 
}
\label{densprof_clash}
\end{figure}

\begin{figure}
\hspace*{-3mm}\includegraphics[width=0.5\textwidth,angle=0.0]{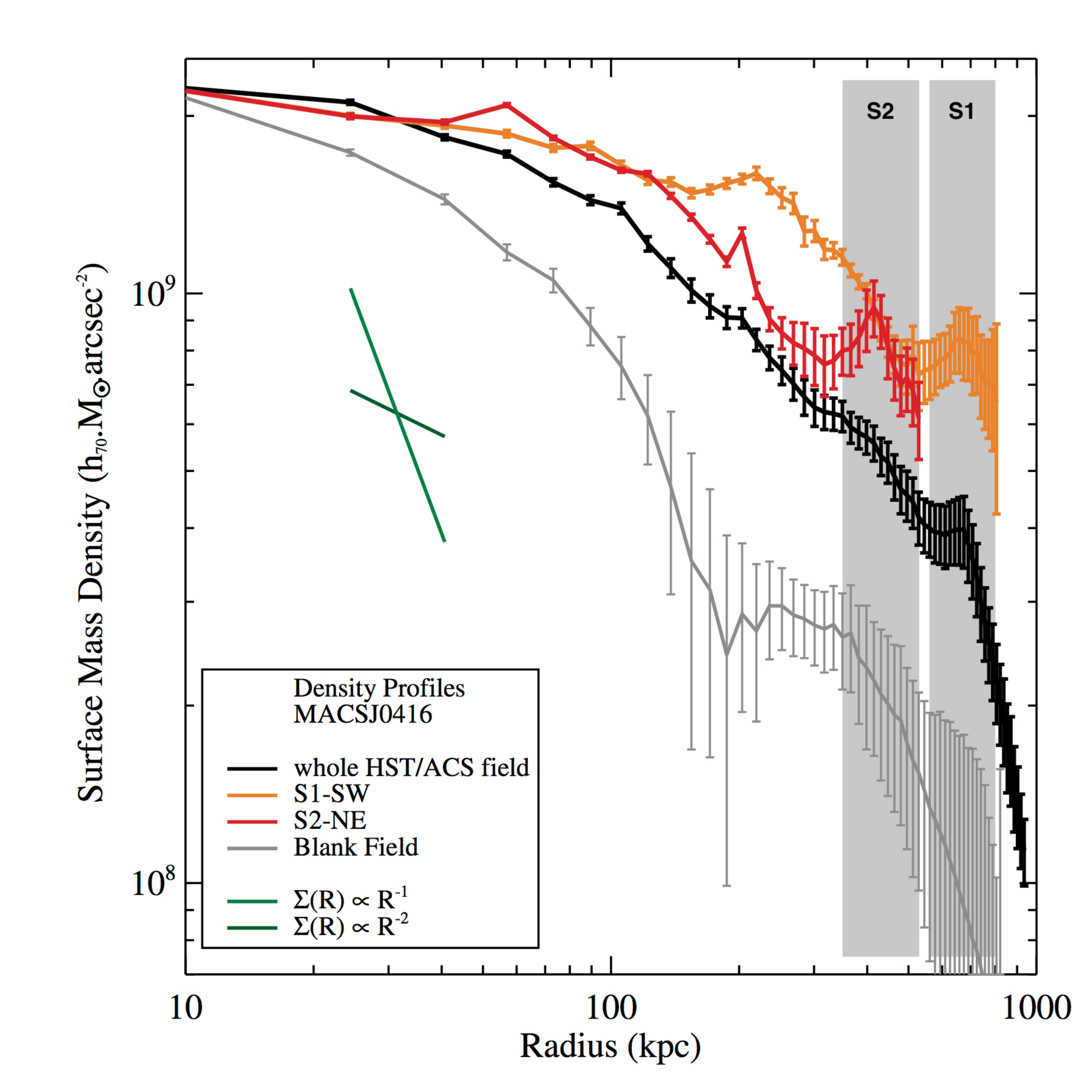}
\caption{Surface density profiles obtained with our complete gravitational lensing analysis (black curve). In orange we show the density profile we obtain in a triangular region designed from the cluster centre and including S1. We show the same density profile in red but for S2.
Finally we designed a triangular region into which there is no apparent sub-structures, and it is shown in grey. The two dark grey shaded areas mark the region of substructures S1 and S2.}
\label{substr_dens}
\end{figure}

Using the grid method described in Sect.~\ref{gridmethod}, we reconstruct the mass distribution of MACSJ0416 within the HFF ACS field of view.  Fig.~\ref{m0416cons} shows contours (white) of the resulting surface mass density overlaid on the ACS colour image.  The mass distribution is highly elongated and features no pronounced substructure at significant distances from the major axis, which is typical of merging clusters. We note as well that the mass distribution is more elliptical than with the SL model. This means that the initial SL model was not only under predicting the mass density, but also the shear in the outskirts. Mirroring the system's morphology in the optical and X-ray regime (magenta and cyan contours in Fig.~\ref{m0416cons}) our lensing reconstruction of the total gravitational mass yields again a strongly bimodal distribution.  In order to measure the radial mass profile, we define the global cluster centre at $\alpha=64.0364$ deg and $\delta=-24.0718$ deg (this is the same centre as used in J14 and marked by a yellow cross in Fig.~\ref{m0416cons}).  As a test of the consistency of our mass reconstruction techniques we can compare the projected mass of $M_{\rm SL} (R{<}320~{\rm kpc}) = (3.26 \pm 0.03)\, \times 10^{14}$ $h_{70}^{-1}$ M$_{\odot}$ measured by J14 using exclusively strong-lensing features with the value of $M_{\rm SL+WL} (R{<}320~{\rm kpc}) = (3.15\pm 0.13)\times 10^{14}$ $h_{70}^{-1}$ M$_{\odot}$ obtained by our joint strong- and weak-lensing analysis. The precision attained here is lower since, at this distance from the cluster core, weak-lensing constraints already contribute to the total mass. Both methods also yield very similar maps of the mass distribution within the cluster core region.

Fig.~\ref{densprof} compares the surface density profiles obtained by this analysis with the ones reported by J14 based on their strong-lensing mass model.  Note the very good agreement at the outer edge of the multiple-image region (shaded area). At larger radii ($R{>}250$ $h_{70}^{-1}$kpc) the predictive power of the model of J14 weakens, due to the lack of 
structure massive enough to induce strong lensing.  By contrast, the combined strong- and weak-lensing mass model remains sensitive to the less pronounced density variations at the outskirts of the cluster.

Pre-HFF strong+weak lensing mass models were also published in \cite{gruen14}, and R14. 
\cite{gruen14} present a weak-lensing analysis of the Wide-Field Imager SZ Cluster of Galaxy (WISCy) sample based on data collected with the Wide-Field Imager \citep[WFI, ][]{baade99} on the 2.2m MPG/ESO telescope at La Silla. For MACSJ0416, \cite{gruen14} obtain mass values that agree well with ours: $M (R{<}200~{\rm kpc}) = 1.40^{+0.22}_{-0.23}\times 10^{14}$ $h_{70}^{-1}$ M$_{\odot}$ (Daniel Gruen private communication) vs.\ our value of $M (R{<}200~{\rm kpc}) = (1.66\pm 0.05)\times 10^{14}$ $h_{70}^{-1}$ M$_{\odot}$.
More recently, \cite{umetsu14} published results from a weak-lensing analysis of the CLASH cluster sample using SUBARU data.  For MACSJ0416, \cite{umetsu14} obtain a mass value of $M (R{<}950~{\rm kpc}) = 0.98 \pm 0.14\times 10^{15}$ $h_{70}^{-1}$ M$_{\odot}$ (Keiichi Umetsu, private communication) that agrees well with our measurement of $M (R{<}950~{\rm kpc}) = 1.15 \pm 0.07\times 10^{15}$ $h_{70}^{-1}$ M$_{\odot}$.  \cite{umetsu14} reconstructed projected mass density profiles for their cluster sample from a joint likelihood analysis of Subaru shear and magnification measurements. The errors include the estimated contribution from uncorrelated large-scale structures projected along the line of sight. Their result is shown in Fig.~\ref{densprof_clash} and found to be in good agreement with the density profile obtained by us for MACSJ0416.
Finally, the pre-HFF model presented by our team in R14, which combines strong- and weak-lensing constraints but uses a parametric approach similar to the one presented in J14, yields $M (R{<}200~{\rm kpc}) = (1.63\pm 0.03)\times 10^{14}$ $h_{70}^{-1}$ M$_{\odot}$.

\subsubsection{Detection of Substructure}
\label{substr_detection}

As mentioned previously, the SL-only results from J14 are in excellent agreement with those presented here from an analysis that combines strong and weak lensing. The advantages of the latter come to bear particularly in the outskirts of the region probed by the HFF images, where any sufficiently massive substructures will reveal their presence by creating weak gravitational shear in the images of background galaxies.

In the case of MACSJ0416, we detect two new substructures at high significance in our mass map; their centres are labeled S1 and S2 in Fig.~\ref{m0416cons}.  The projected masses of S1 and S2, estimated within an aperture of $\sim$100~kpc, are $(4.22\pm0.56)$ and $(1.5\pm 0.20) \times 10^{13}$ $h_{70}^{-1}$ M$_{\odot}$, respectively, values typical of galaxy groups.  A tentative identification of S1 as a group of galaxies is supported by the presence of a coinciding galaxy overdensity, clearly visible also as a peak in the cluster light distribution (Fig.~\ref{m0416cons}). For S2, the association with a galaxy group is much less obvious. Physical characteristics of S1 and S2 are listed in Table~\ref{tab_substr}. The distribution of cluster light as shown in Fig.~\ref{m0416cons} is obtained by smoothing a map of the flux in the ACS-F814W band from cluster members with a Gaussian kernel ($\sigma{=}9\arcsec$). 

The imprint of these minor mass concentrations on the overall mass density profile can be seen in Fig.~\ref{densprof} in the form of enhancements at $\sim$450 $h_{70}^{-1}$kpc (S2), and (much more clearly) at $\sim$650$h_{70}^{-1}$kpc (S1) from the cluster centre.
In order to test whether these features are indeed caused by S1 and S2, we follow the same procedure as used in J12 to confirm the detection of the large-scale filament in the field of MACSJ0717.5+3745 and define three triangular regions, anchored at the global cluster centre and extending toward S1, S2, and (for control purposes) a region towards the NW of the field centre that is void of any mass overdensities. Fig.~\ref{substr_dens} presents the resulting surface density profiles in these three regions and shows indeed that the excess surface mass density can be attributed to S1 and S2, while no significant variations are observed in the radial surface mass density profile of the control field.

\begin{table}
\begin{center}
\begin{tabular}[h!]{cccccc}
\hline
\hline
\noalign{\smallskip}
$ID$ & R.A.\ (deg)& Dec.\ (deg) & $M (10^{13} h_{70}^{-1}$M$_{\odot})$ & $\sigma$ & $D_{C-S} (kpc)$ \\
\hline
S1 & 64.016542 & -24.094906 & $4.22 \pm 0.56$ &  7.5 & 580 \\
S2 & 64.06097 & -24.063636 & $1.46 \pm 0.20$ & 7.3 & 470 \\
\noalign{\smallskip}
\hline
\hline
\end{tabular}
\caption{Coordinates, mass within a $\sim$100~kpc aperture, significance of detection, and distance to the cluster centre ($D_{C-S}$) for the two substructures detected in the outskirts of MACSJ0416.
}
\label{tab_substr}
\end{center}
\end{table}

\subsection{Distribution of Stellar Mass}
\label{stellar_distri}
To measure the stellar mass distribution, $M_{\ast}$, across the ACS field of view, we use the same method as in J12. We compute the relation $\log(M_{\ast}/L_{\rm K}) = a z + b$, established by \cite{arnouts07} for quiescent (red) galaxies in the VVDS sample \citep[][]{lefevre05}, and adopt a Salpeter initial mass function (IMF). Here $L_{\rm K}$ is the galaxy's luminosity in the K-band, $z$ is its redshift, and the parameters $a$ and $b$ are given by:
\begin{eqnarray}
 a & = & -0.18 \pm 0.03, \nonumber \\
 b & = & -0.05 \pm 0.03. \nonumber
 \end{eqnarray}

We apply this relation to our catalogue of 146 cluster members used in our mass model (see Sect.~\ref{CM_model}).  We estimate the K-band luminosity of cluster members observed in the F814W band using theoretical 
models from \cite{BC03} to predict the typical $(m_{\rm F814W}-m_{\rm K})$ colours. We assume a passively 
evolved galaxy observed at $z=0.4$, with a range of exponentially decaying star-formation histories within the range $\tau=0.1-2$ Gyr. This provides a typical colour $m_{\rm F814W}-m_{\rm K} =1.14\pm 0.04$ (AB system). 
Using the public data obtained from GSAOI observations in $K_s$ band \citep{schirmer14}, we confirmed this colour for cluster member galaxies located in the central 100x100 arcmin$^{2}$ of the cluster.
The resulting projected mass density in stars decreases in proportion to the total projected mass density depicted in Fig.~\ref{densprof}. We measure a mass-to-light ratio across the study area of M$_{\ast}$/L$_{\rm K} = 0.99 \pm 0.03$ M$_{\odot}$/L$_\odot$.
To compare our results with those obtained by \cite{leauthaud12b} for COSMOS data, we need to adjust our measurements to account for the different IMF used by these authors. Applying a shift of 0.25 dex to our masses to convert from a Salpeter IMF to a Chabrier IMF, we find $(M_{\ast}/L_{\rm K})_{\rm Chabrier} = 0.78 \pm 0.02$ M$_{\odot}$/L$_\odot$ for quiescent galaxies at $z\sim0.4$, in good agreement with \cite{leauthaud12b}.

The fraction of the total mass in stars, $f_{\ast}$, {\sl i.e.}, the ratio between the stellar mass and the total mass of the cluster derived from our lensing analysis within the ACS field of view, is $f_{\ast} = 3.15 \pm 0.57\%$ (assuming a Salpeter IMF). The latter value is slightly higher than the one derived by \cite{leauthaud12b}. The difference might be due to different limiting K-band magnitudes or differences in the galaxy environments probed (the COSMOS study was conducted for groups with a halo masses between 10$^{11}$ and 10$^{14}$ $h_{70}^{-1}$ M$_{\odot}$, and extrapolated to halos of ${\sim}10^{15}$ $h_{70}^{-1}$ M$_{\odot}$). Another cause might be the use of the analytical \cite{arnouts07} relation to estimate the stellar mass. As discussed by \cite{ilbert10}, the M$_{\ast}$/L$_{\rm K}$ relation used here is only calibrated for massive galaxies, while in practice this ratio varies with galaxy age and colour. Therefore, the \cite{arnouts07} relation overestimates the stellar masses of low-mass galaxies. Although we do not expect our cluster member sample to be dominated by low-mass galaxies, a bias cannot be firmly ruled out. Finally we compute the total stellar mass within our study area and find M$_{\ast} = (3.10 \pm 0.01)\times 10^{13}$ $h_{70}^{-1}$ M$_{\odot}$.
Upcoming HFF data in F160W will provide more direct estimates of the stellar masses.

\subsection{Intra-Cluster Medium}
\label{gas_distri}

\subsubsection{X-ray Morphology}
\label{Xraymorph}
Fig.~\ref{m0416cons} shows  the X-ray contours (in cyan) of the adaptively smoothed X-ray emission as observed with  \emph{Chandra}, as described in Sect.~\ref{chandra_obs}. The X-ray emission shows two peaks (labelled X1 and X2 in Fig.\ref{m0416cons}), located at R.A.${=}$64.038458 deg, Dec${=}-$24.067361 deg (X1) and R.A.=64.029792 deg, Dec${=}-$24.08025 deg (X2), and exhibits a strong elongation in the NE-SW direction. The main peak of the X-ray emission (X1) coincides with the first mass concentration detected in the cluster core by our lensing analysis (C1 in Table~\ref{table_SLparam}). However, the second peak (X2)  located ${\sim}45\arcsec$ (${\sim}250$ kpc) SW of X1, does not coincide with the mass concentration C2 in Table~\ref{table_SLparam} and is offseted to the SW by ${\sim}15\arcsec$. In the following we use several models to characterise the gas distribution that gives rise to the X-ray morphology of MACSJ0416.

Acknowledging that the X-ray emission from MACSJ0416 is clearly neither unimodal nor spherically symmetric, we attempt to model the observed X-ray surface brightness distribution as the superposition of two elliptical $\beta$-models in \emph{Sherpa}, leaving the position of the centroids free to vary. This model provides a good description of the data and yields best-fit centroids for the two components of R.A.${=}$64.040604 deg, Dec${=}-$24.06654 deg and R.A.$=$64.029713 deg, Dec${=}-$24.081072 deg, respectively. The $1\sigma$ uncertainty of these positions is approximately 2$\arcsec$.  The model returns best-fit values of $r_{c1}=152\pm24$ kpc and $r_{c2}=68\pm17$ kpc for the core radii of C1 and C2, respectively. The centroid of the main component is thus slightly shifted (by ${\sim}12\arcsec$) to the NE of the X-ray peak, which is not surprising given the irregular morphology of the X-ray emission. The centroid of the second model component, however, coincides with the X-ray surface-brightness peak X2. 

Proceeding to less massive structures identified in our reconstruction of the mass distribution in MACSJ0416 (Fig.~\ref{m0416cons}), we also search for X-ray emission from substructure S1, tentatively  identified as a galaxy group in Sect.~\ref{substr_detection}.  No evidence of X-ray emission from S1 is discernible in Fig. \ref{m0416cons}.  In order to obtain a quantitative assessment of the X-ray luminosity and thus gas mass of S1, we add a third, group-sized component ($r_{\rm c}=100$ kpc, $\beta=0.7$) to our model, tied to the position of S1 as defined in Table~\ref{tab_substr}. We find that the data do not require this third component.  The upper limit to the 0.7--7 keV photon flux of $3.2\times10^{-6}$ cm$^{-2}$ s$^{-1}$ within a 20$\arcsec$ radius (90\% confidence level) corresponds to an upper limit to the X-ray luminosity and gas mass of mass concentration S1 of L$_{\rm X}<6.2\times10^{42}$ erg s$^{-1}$ (unabsorbed, 0.1--2.4 keV) and M$_{\rm gas}({<}20\arcsec)<2.9\times10^{11}$ $h_{70}^{-1}$ M$_\odot$, respectively.

\subsubsection{Spectral X-ray Analysis}
\label{Xrayspec}

\begin{figure}
\hspace*{-2mm}\includegraphics[width=0.5\textwidth,angle=0.0]{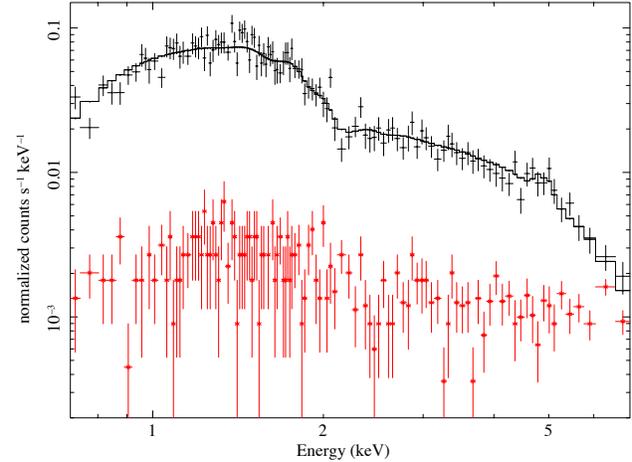}
\caption{Global \empty{Chandra}/ACIS-I spectrum of the cluster within 2 arcmin radius around the X-ray peak. The solid line shows the best-fit single-temperature APEC model. The red data points indicate the background level as estimated from a source-free region.}
\label{fig:xrayspec}
\end{figure}

In order to further constrain fundamental properties of the intra-cluster medium (ICM) of MACSJ0416, we examined the X-ray spectrum of the diffuse cluster emission within a radius of 2 arcmin from the primary X-ray peak (see Fig. \ref{fig:xrayspec}). Obvious point sources were excised from the event file prior to the spectral extraction, and a source-free region was defined within the same ACIS-I chip to estimate the local background. We modeled the spectrum with a single-temperature APEC model \citep[][]{smith01xray} absorbed by the Galactic hydrogen column density, which we fixed at the 21cm value of $N_H=3.05\times10^{20}$ cm$^{-2}$ \citep[][]{kalberla05}. Because of the poor photon statistics of the archival ACIS-I observations (${\sim}4500$ source counts in the 0.7--7 keV band), we grouped the spectral channels to obtain a minimum of 20 counts per bin and used the C-statistic \citep[][]{cash79} for the fitting procedure. We obtained an average cluster temperature of $11.0_{-1.3}^{+1.4}$ keV  and an Fe abundance of $0.20_{-0.08}^{+0.09}Z_\odot$, where the quoted uncertainties represent the 1$\sigma$ confidence level. According to the $M$-$T$ relation of \cite{arnaud05}, this temperature corresponds to a total mass of M$_{\rm 2500}=(4.8_{-0.7}^{+0.9})\times10^{14}\, $ $h_{70}^{-1}$ M$_\odot$.  With R$_{\rm 2500}$ corresponding roughly to 400~kpc, this result is consistent with the value of M$(R{<}400\,{\rm kpc})=(4.12\pm 0.17)\times10^{14}\, h_{70}^{-1}$M$_\odot$ measured by our lensing analysis. Most studies \citep[e.g.,][]{nagai07,rasia12,nelson12} predict that the lack of thermalization of the gas in violent cluster mergers should lead to an underestimation of the X-ray mass compared to the true mass. However, according to our analysis MACSJ0416 lies on the $M$-$T$ relation, in spite of ongoing merger activity. While not of great significance in its own right, this result agrees with the low scatter observed around that relation in cluster samples \citep[e.g.,][]{mahdavi13}. This is important for future X-ray surveys (e.g., \emph{eROSITA}), which will use the X-ray temperature as a proxy for cluster mass.

A region of special interest is the core of the NE cluster which appears very compact in Fig.~\ref{m0416cons} and perfectly aligned with the associated Brightest Cluster Galaxy (BCG).  To test the hypothesis that this cluster component might host a cool core, we extracted the X-ray spectrum of the two components of MACSJ0416 separately.  We measure k$T{=} 10.3_{-0.8}^{+1.1}$ keV and k$T{=} 13.6_{-1.9}^{+2.2}$ keV for the NE and SW sub cluster, respectively.  Attempts to directly fit an isothermal plasma model to the current archival data within a circle of 10$\arcsec$ radius of X1 yield unphysical results of either extreme excess absorption of several 10$^{21}$ cm$^{-2}$ (equivalent column density of neutral hydrogen) or temperatures well over 20 keV, i.e., far outside the range that can be constrained with Chandra. Given the poor photon statistics (less than 700 net photons) we do not take these results at face value but rather as indication that an isothermal model is inappropriate. Although we have thus currently no direct spectroscopic evidence of a cool core of the NE cluster component, the data appear to suggest the presence of multi-phase gas in this region.  

\subsubsection{Gas Density and Gas Mass}
We follow the procedure described in \cite{eckert12} to estimate the three-dimensional gas-density and gas-mass profiles of MACSJ0416. Accounting for vignetting effects, we extracted a surface-brightness profile for a set of concentric annuli of 5$\arcsec$ width centered on the primary X-ray peak, and estimated the local background at radii beyond 4$\arcmin$. Cluster emission is detected out to $\sim3\arcmin$ ($\sim$1 Mpc). The resulting profile was deprojected assuming spherical symmetry using the method of \cite{kriss83}. We converted the deprojected profile into an emission-measure profile assuming a constant temperature of 7.8 keV (see above), and inferred the gas-density profile by assuming constant density within each radial shell. Finally, the gas-mass profile was calculated by integrating the gas-density profile in concentric shells.

We measure gas masses of M$_{\rm gas}(R{<}500~{\rm kpc})=(3.4\pm0.2)\times10^{13}$ $h_{70}^{-1}$ M$_\odot$ and M$_{\rm gas}(R{<}1~{\rm Mpc})=(8.6\pm0.7)\times10^{13}$ $h_{70}^{-1}$ M$_\odot$. In this context, a note is in order regarding systematic effects. Given the irregular morphology of the cluster, the assumption of spherical symmetry might lead to an incorrect gas mass. However, since the morphology of the system beyond the inner regions appears relatively regular, only the gas masses observed in the central regions are significantly affected. Moreover, as stated in \cite{rasia11}, gas-mass measurements are relatively unaffected by the presence of merging substructures, due to the quadratic dependence of the X-ray emissivity on gas density.  We thus expect little systematic bias in our measurement, in spite of the unrelaxed morphology of the system.

\subsection{Baryon Fraction}
\label{bar_distri}

\begin{figure}
\hspace*{-2mm}\includegraphics[width=0.5\textwidth,angle=0.0]{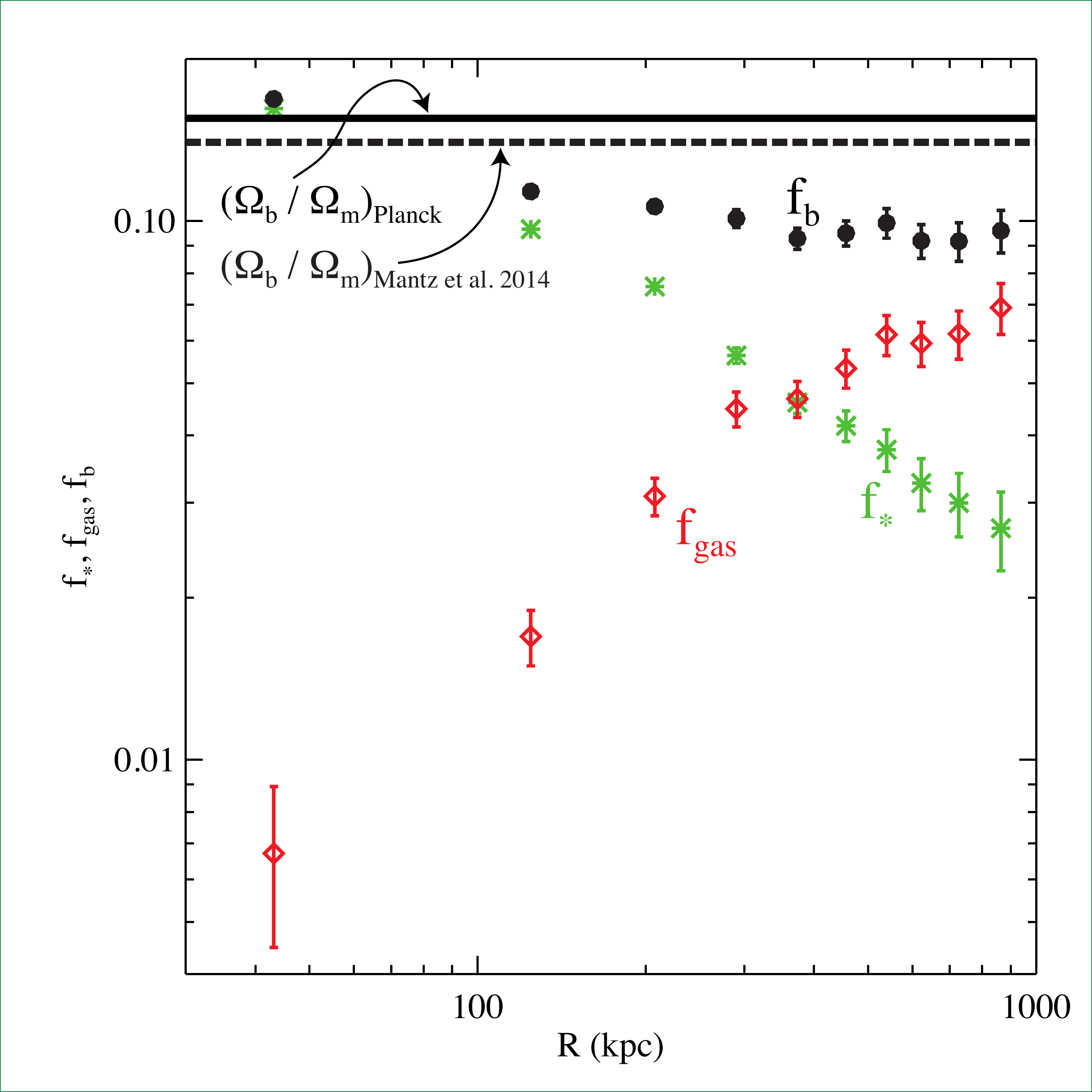}
\caption{Fraction of stars, $f_{\ast}$, of gas, $f_{gas}$, and fraction of baryons, $f_{b}$ present in MACSJ0416. We also plot the baryon fraction measured by Planck ($f_{b} = \Omega_b/\Omega_m = 0.1551 \pm 0.0055$), thick black line, and the one measured by \citet{mantz14} ($f_{b} = \Omega_b/\Omega_m = 0.14 \pm 0.02$), thick dashed black line.}
\label{baryons_frac}
\end{figure}

Fig.~\ref{baryons_frac} shows the fraction of baryons in stars (green asterisks, see Sect.~\ref{stellar_distri}) and gas (red diamonds, see Sect. \ref{gas_distri}) as a function of cluster-centric radius.  Both profiles exhibit trends typical for massive galaxy clusters, in which energy input from cluster mergers as well as feedback from active galactic nuclei \citep{nulsen05} and galactic winds \citep{metzler94} raise the entropy of the intracluster medium. As a result, the hot gaseous atmosphere expands, and the gas fraction increases significantly with radius \citep[e.g.][]{ettori99,vikhlinin06}. Conversely, the stars condensate within the massive central galaxies, and the stellar fraction shows the opposite trend. 

The total baryon fraction is shown by orange circles in Fig.~\ref{baryons_frac}.
Within 1 Mpc from the cluster core, we measure a gas fraction of $0.072\pm0.007$ and a stellar fraction of $0.027\pm0.004$. The total baryon fraction within this aperture is thus $f_{\rm bar}=0.099\pm0.008$. This value is $5\sigma$ below the cosmic baryon fraction measured by \cite{planck13_cosmo}, $f_{b} = \Omega_b/\Omega_m = 0.1551 \pm 0.0055$ (black thick line in Fig.~\ref{baryons_frac}), and also discrepant with the cluster measurement of \citet{mantz14}, $f_{b} = \Omega_b/\Omega_m = 0.14 \pm 0.02$, at more than $4\sigma$ confidence (dashed black thick line in Fig.~\ref{baryons_frac}). 

This tension might be due to several factors. First, our analysis may have missed a significant fraction of the total stellar mass of the cluster, since the selection of cluster members considers mainly red galaxies. The contribution from less massive (and fainter) star-forming cluster galaxies is not taken into account. In addition, intra-cluster light (ICL) can account for 10--40\% of the total stellar mass \citep[e.g.][]{gonzalez07,giodini09,lagana13}. However, we estimate that, overall, the missing stellar content can contribute at most to 1\% of the total cluster mass. Another explanation is that a large fraction of the baryons resides outside the region sampled in our study. Recent studies \citep[e.g.][]{simionescu11,eckert13b} have shown that the hot gas fraction continues to increase beyond $R_{500}$ and eventually reaches the universal baryon fraction. It is likely that a significant fraction of the baryons indeed resides beyond 1 Mpc from the cluster centre, although the gas fraction of 7\% measured here is still significantly lower than the typical values measured in massive clusters around $R_{500}\sim 400~kpc$ \citep[$f_{\rm gas}\sim0.13$,][]{vikhlinin06,pratt09}. Finally, the total mass used to derive the baryon fraction could be overestimated by substructure along the line of sight. Our dynamical analysis of the member galaxies indeed revealed a difference of 800 km s$^{-1}$ between C1 and C2, and the substructure S1 appears to be largely aligned with our line of sight (see Sect. \ref{radvel}). Since all of the aforementioned biases are known to be present but difficult to account for, we suggest that the deficit of baryons observed in Fig.~\ref{baryons_frac} is probably due to a combination of these effects. In particular, it is likely that because of the merging activity a significant fraction of the baryons reside in the outskirts of the cluster, and given the presence of significant line-of-sight structure the total lensing mass is also likely overestimated by some fraction. \cite{BK11} demonstrated that up to $\sim$20\% scatter can be expected for weak-lensing mass measurements in the case of massive galaxy clusters, due to the presence of correlated and uncorrelated large-scale structures. The impact of these effects is, however, largely limited to large cluster-centric distances ($>$3~Mpc), well beyond the value of $\sim$1~Mpc from the cluster centre to which we map the mass distribution of MACSJ0416 in this study.

\subsection{Radial Velocities of Cluster Galaxies}

As reported in Sect.~\ref{totalmass}, the mass map derived by our joint strong- and weak-lensing analysis reveals four significant mass concentrations:  the two main merger components, as well as two smaller components (labeled S1 and S2 in Fig.~\ref{m0416cons}) detected at more than 7$\sigma$ confidence, that are located about 500~kpc (in projection) from the overall centre of the cluster. 

\begin{figure}
\hspace*{-2mm}\includegraphics[width=0.5\textwidth,angle=0.0]{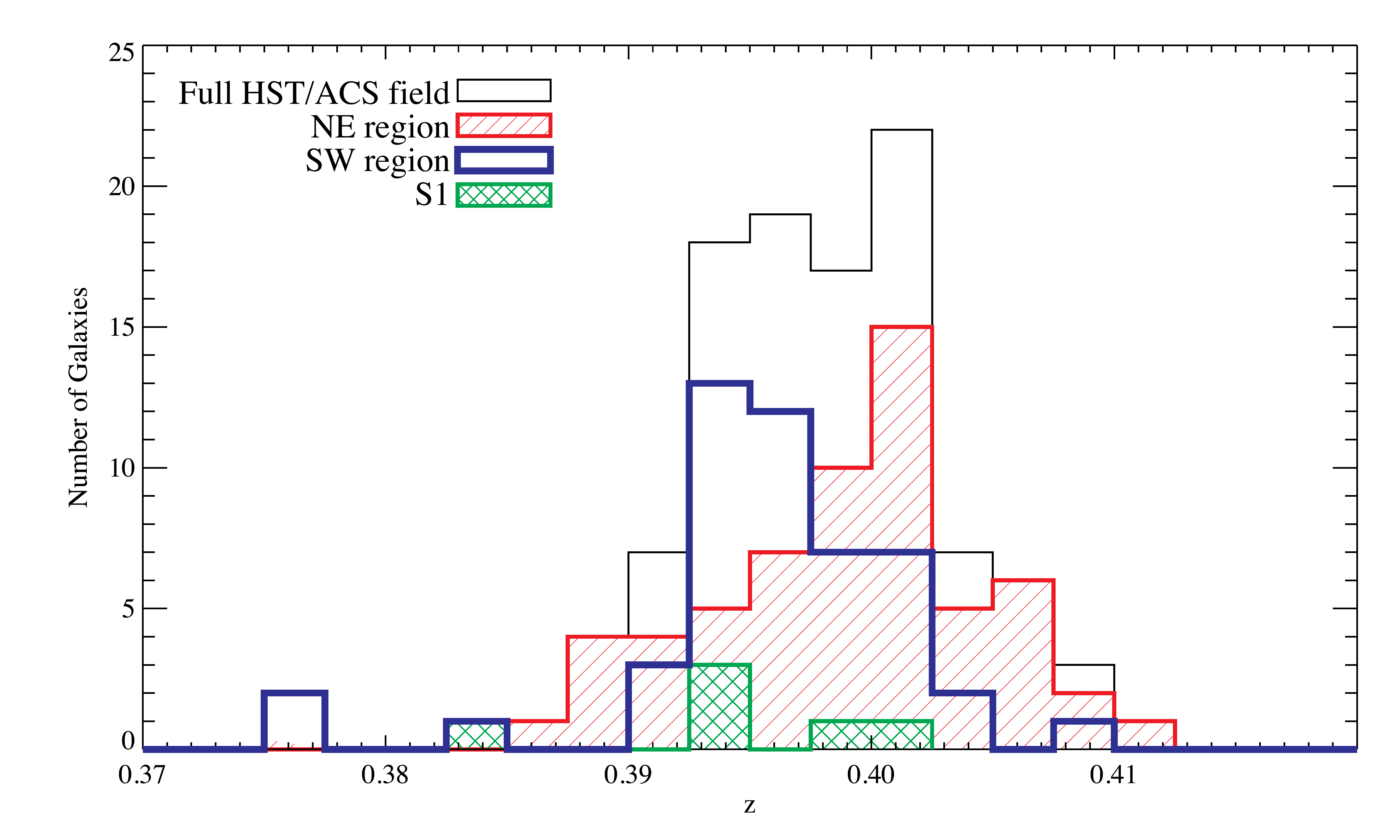}
\caption{Distribution of spectroscopic galaxy redshifts in the field of MACSJ0416. The red and blue histograms show the different redshift distributions in the NE and SW sections of the ACS field, respectively, the green histogram shows the redshift distribution within the substructure S1 (see text for details).}
\label{spectro_merg}
\end{figure}

As part of our attempt to clarify the nature and role of all four of these components in the assembly of MACSJ0416 (Sect.~\ref{discussion_dynamic}) we examine the redshift distribution of galaxies in the respective regions using the spectroscopic data described in Sect.~\ref{zspec_zphot_obs}.  Using the {\sc Rostat} package \citep{beers} we find an overall redshift of 0.3980 for MACSJ0416 and a global velocity dispersion of 740 km s$^{-1}$, based on 106 spectroscopic redshifts. For a first global assessment, we divide the field of view along a boundary that runs perpendicular to the line connecting the main NE and SW cluster components (the apparent projected merger axis), intersecting it at its midpoint.  Fig.~\ref{spectro_merg} shows the overall distribution of spectroscopic galaxy redshifts (black), as well as, separately, the redshift distributions for the NE (red) and SW regions (blue) thus created.  A two-sided Kolmogorov-Smirnov test yields a probability of less than 0.5\% for the hypothesis that the redshift distributions for the NE and SW regions are drawn from the same parent population.  Their average redshifts are 0.3990 (NE) and 0.3966 (SW). The difference becomes more pronounced when the modes of the two redshift histograms are considered: 0.4013 (NE) and 0.3938 (SW).

Proceeding to the much less massive structures S1 and S2, we measure an average redshift of $z_{\rm S1} = 0.3944$ from six cluster members with spectroscopic redshifts within 13$\arcsec$ ($\sim$70 kpc), but are unable to estimate the redshift of S2 as we presently have no spectroscopic redshift for galaxies even within  $15\arcsec$ of its centre.

\label{radvel}

\section{The Merger Geometry and History of MACSJ0416.1$-$2403}
In the following we combine our findings regarding the distribution of total gravitational mass, ICM, and cluster galaxies to derive a self-consistent picture of the evolutionary state and three-dimensional merger history of MACSJ0416. 

\label{discussion_dynamic}
\subsection{Evidence}
The bimodal distribution of galaxy redshifts shown in Fig.~\ref{spectro_merg} and a projected separation of the two components of less than 300 kpc identify MACSJ0416 unambiguously as an active merger.  The geometry of the collision, or even the answer to the question whether we observe MACSJ0416 before or after core passage, are not immediately obvious though.

Since MACSJ0416 has decoupled from the Hubble flow, the difference between the redshift distributions of galaxies in the NE and SW part of the cluster (Sect.~\ref{radvel}) can safely be attributed to peculiar velocities.  The difference between the means (modes) of the distributions corresponds to 500 (1600) km s$^{-1}$, while the difference in redshift between the BCGs of C1 and C2 (for simplicity we here adopt the nomenclatures of Table~\ref{table_SLparam}) implies a relative velocity of over 800 km s$^{-1}$ --- all remarkably high values compared to the global velocity dispersion of the system of 750 km s$^{-1}$.   With peak collision velocities in massive mergers typically ranging from 1000 to 3000 km s$^{-1}$, the high relative radial velocity of the two subclusters strongly suggests a merger axis that falls markedly outside the plane of the sky. In a relative sense, C2 is thus moving toward us, while C1 is receding. 

For an assessment of the direction of motion of the components of MACSJ0416 in the plane of the sky, as well as of the three-dimensional merger geometry in general, an inspection of the relative offsets (if any) between the collisional and non-collisional cluster components (the intra-cluster gas, and galaxies as well as dark matter) proves instructive. Binary head-on mergers \citep[BHOM;][]{ME12} will feature a binary X-ray morphology before the collision, and a more unimodal morphology after (unless the merger axis falls very close to our line of sight).  Offsets between gas and galaxies (and dark matter) will increase throughout the collision, as the non-collisional cluster components proceed unimpeded while the viscous ICM is shocked and slowed during the collision. Regardless of the merger axis, these offsets would be apparent in both participants in a BHOM. This is not the case for MACSJ0416.  As shown in Sect. \ref{Xraymorph}, collisional (gas) and non-collisional matter (galaxies and dark matter) coincide well for the NE component (C1 and X1), but are clearly displaced from each other (at 7$\sigma$ significance) for the SW component (C2 and X2). Since non-collisional matter has to lead collisional matter in a merger, we infer that C2 is moving toward C1 (in projection). 

An intriguing final piece of evidence is provided by the non-detection of structure S1 by \emph{Chandra}.  The lensing mass of S1 of $M_{\rm S1} (R{<}110\, {\rm kpc})=(4.22\pm 0.56)\times 10^{13}$ $h_{70}^{-1}$ M$_\odot$ (Table~\ref{tab_substr}) is substantial, and yet the mass derived for it from the upper limit to its X-ray luminosity of $L_{\rm X}{<}3.9\times10^{42}$ erg s$^{-1}$ is $M_{2500}<10^{13} h_{70}^{-1}$ M$_\odot$, based on the $L_{\rm X}$--$M$ relation for galaxy groups \citep{sun09,eckmiller11}. This apparent conflict, as well as the absence of hot gas in S1 as evinced by our tight upper limit to its gas fraction of $f_{\rm gas}<0.007$ (see Sect.~\ref{Xraymorph}), are easily explained though if S1 is in fact part of an unvirialised filamentary structure, almost aligned with our line of sight.

\subsection{Tentative merger scenario}

From the evidence compiled in the preceding section we conclude that MACSJ0416 is (a) not a BHOM, i.e., the merger proceeds at significant impact parameter, rather than head-on, and (b) that the merger axis is greatly inclined with respect to the plane of the sky.   MACSJ0416 is thus reminiscent of the merging system MACSJ0358.8$-$2955 for which \citet{hsu13} conclude that the lower-mass component is likely moving along a trajectory that is curved towards our line of sight. The morphology of MACSJ0416 is very similar, except that the offset observed between the gas and the dark matter in the SW component (X2 and C2) goes in the opposite way as seen in MACSJ0358.8$-$2955. 

We propose two alternative scenarios for the merger history of MACSJ0416, both of which are consistent with the present three-dimensional geometry as outlined above\footnote{Both of our scenarios are also in qualitative agreement with the one advanced by \cite{diego14} who propose that MACSJ0416 is merging along an axis that is only mildly inclined with respect to our line of sight.}.

\begin{figure*}
\includegraphics[width=\textwidth]{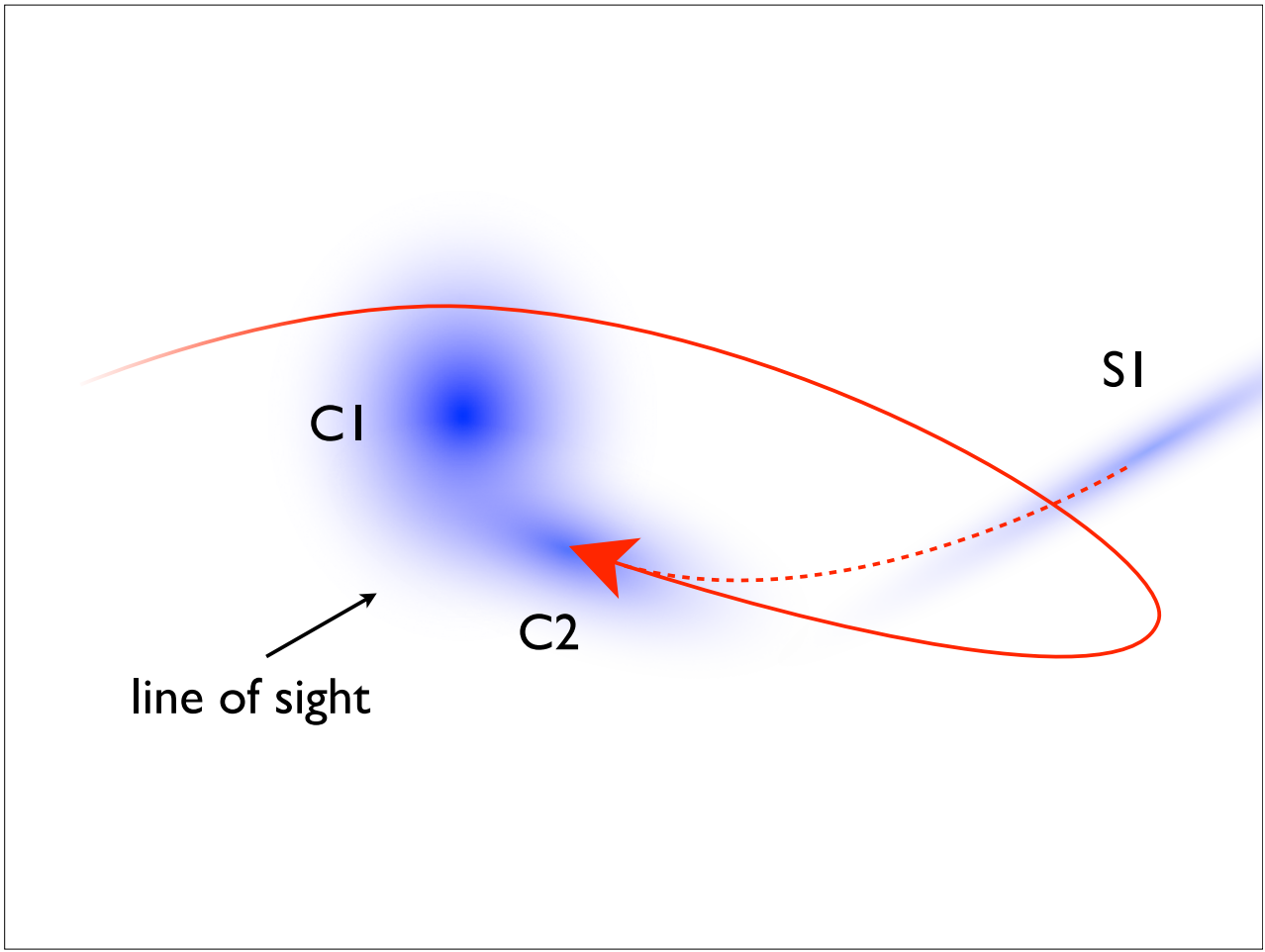}
\caption{Schematic sketch of our hypothesised merger scenarios in a face-on view of the plane of the collision.  The dotted red line marks the trajectory of C2 in our first scenario, in which the system is observed before its first core passage. The solid red line represents the second (preferred) scenario, in which C2 has passed C1 and is viewed after turn-around as it approaches the more massive cluster component for the second time.  In this second scenario, filament S1 plays no significant role in the ongoing merger and may be positioned well before the cluster (along our line of sight), outside the area covered by this sketch.}
\label{sketch}
\end{figure*}

 \emph{Scenario \#1}: The SW component C2, observed near core passage, moves along a curved trajectory that originates in a large-scale filament, part of which is detected as substructure S1 by our lensing reconstruction of the mass distribution.  Like for C2 itself, the mean redshift of galaxies near S1 of $z_{\rm S1}{=}0.3944$ is lower than that of the NE cluster C1, implying a radial velocity toward the observer of about 1000 km s$^{-1}$.  The SW region of Fig.~\ref{m0416cons} contains a superposition of gas, galaxies, and dark matter from both this putative filament and C2, and hence the differences between the contours of gravitational mass, X-ray surface brightness, and cluster light reflect not only the complex geometry and dynamical history of this system, but are also partly the result of the fact that lensing mass reconstructions collapse the mass of structures along the entire line of sight, virialised or not.  Since, in this scenario, the merger of C1 and C2 resembles a "fly-by" at the time of observation, the trajectory of the SW component has, so far, only grazed the core of the NE cluster which remains largely undisturbed, while ram pressure causes a disassociation between the ICM (X2) and the collisionless constituents (C2) of the approaching cluster (as seen in Fig.~\ref{m0416cons}).  A sketch of this scenario and of the trajectory of C2 is shown in Fig.~\ref{sketch} (dotted line) in a face-on view of the orbital plane.
  
 \emph{Scenario \#2}: The SW component C2 approaches C1 just like in  \emph{Scenario \#1}, but it does so for the second time.  In this scenario, C2 does not originate in the large-scale filament S1; rather, C2 originally fell toward C1 from the opposite direction, passing the cluster core at a significant distance before turning around for its second approach.  The core of C1 has been significantly disturbed by the earlier passage of C2, but any gas sloshing or shock heating induced by this interaction is almost imperceptible from our viewing angle, as the resulting cold fronts or shock fronts are not viewed edge-on, but are mainly projected onto the core of C1.  The trajectory of C2 in this scenario is shown by the solid line in Fig.~\ref{sketch}.  In  \emph{Scenario \#2}, filament S1 may be just behind the cluster (along our line of sight), as indicated in Fig.~\ref{sketch}, or far in front of MACSJ0416, well outside the virial regime.  The X-ray evidence of multiphase gas in the core region of component C1 (Section~\ref{Xrayspec}) supports Scenario \#2 which we thus presently favour.
 
More speculatively, and based largely on the apparent lack of a well formed X-ray core, we further propose that the SW component of MACSJ0416 is undergoing its own (minor) merger event.  This second merger in the MACSJ0416 system proceeds again at high inclination with respect to the plane of the sky, is in the post-collision phase, and greatly disturbed the ICM, leading to the (unresolved) flat profile around X2.  While this potential merger within the SW component of MACSJ0416 can be accommodated by either of our merger scenarios, it would fit more naturally into \emph{Scenario \#2} in which tidal forces during the first core passage may have aided the disruption of C2.


Although the currently available data do not allow us to clearly distinguish between these speculative scenarios, the tantalising evidence of shock-heated gas in component C1 (Section~\ref{Xrayspec}) strongly supports \emph{Scenario \#2}.
An opportunity to discriminate between \emph{Scenarios \#1} and \emph{\#2} will be provided by upcoming deep \emph{Chandra}/ACIS-I observations of MACSJ0416. Ultimately reaching a cumulative exposure time of over 300 ks, these observations will be able to detect the presence of shocked gas near X1, as well as between X1 and C2 (clear sign of a previous interaction of the two components), as well as of cold fronts created by sloshing gas near X1. Dramatically improved photon statistics will yield ICM density and temperature maps also around X2 and stand to reveal the true dynamical history of this complex system.

\section{Summary \& Conclusion}

We present a combined X-ray and optical analysis of the massive galaxy cluster MACSJ0416. Using the deep, high-resolution imaging data obtained by the HFF initiative, we model the mass distribution of the cluster over the HST/ACS field of view using a grid-based method that combines both strong- and weak-lensing constraints.  This lensing analysis is complemented by a study of the diffuse intra-cluster medium, based on archival (16ks + 37ks) \emph{Chandra} ACIS-I observation.  Finally, we probe the distribution of mass along the line of sight using spectroscopic redshifts of 106 cluster members. We measure the following global properties:  an overall galaxy velocity dispersion of 741 km s$^{-1}$ and compelling evidence of bulk motions of $\sim$1000 km s$^{-1}$ along the line of sight; a total mass of $M(R{<}950\,{\rm kpc}) = (1.15\pm 0.07)\times 10^{15}$ $h_{70}^{-1}$ M$_{\odot}$; an average mass-to-light ratio of $M_{\ast}/L_{\rm K} = 0.99\pm 0.03$; a gas mass of $M_{\rm gas} (R{<}1\,{\rm Mpc}) = (8.6\pm 0.7)\times 10^{13}$ $h_{70}^{-1}$ M$_{\odot}$; an ICM temperature of k$T=11.0^{+1.4}_{-1.3}$ keV; an Fe abundance in the ICM of 0.20$^{+0.09}_{-0.08}$ Z$_\odot$; and a baryon fraction of $f_{\rm b}(R{<}1\, {\rm Mpc})  = 0.099\pm 0.008$ (5$\sigma$ below the cosmological value estimated by the \emph{Planck} mission).  Importantly, our multi-wavelength study also constrains the spatial distribution of dark and luminous matter in MACSJ0416 and reveals the presence of a massive ($M=(4.22\pm 0.56)\times 10^{13}$ $h_{70}^{-1}$ M$_{\odot}$) but X-ray dark structure that we associate with a line-of-sight filament.

Using all observational evidence, we attempt to unravel the dynamical state and merger history of MACSJ0416. Central to our interpretation is the large offset in radial velocity between the two main cluster components, the fact that only the SW component shows a clear offset between collisional and non-collisional matter, and -- possibly --  our discovery of the aforementioned putative line-of-sight filament.  We propose two alternative merger scenarios, the general geometry of which resembles that advanced by \citet{hsu13} for MACSJ0358.8$-$2955. Our two scenarios for MACSJ0416 differ from one another  primarily with regard to the pre- or post-collision state of the system.  In either case, we assume that the NE component of MACSJ0416 is the more massive one. In our first merger scenario, the trajectory of the approaching SW component is slingshot-like, possibly originating in the aforementioned filament and passing the NE component at a significant impact parameter during the imminent first core passage.  In our second scenario, the referred-to filament plays no role in the merger, which is much farther advanced than in our first scenario and in fact observed after turnaround of the SW component which now approaches the dominant NE component for the second time.  The first passage significantly disturbed the core of the NE component, triggering gas sloshing of a modest cool core and shock heating of the gaseous ambient ICM.  Since our spectral analysis of the ICM in the core region of the NE component finds tentative evidence of multi-phase gas, we currently favour this second scenario. 
Traveling along a vector that is highly inclined with respect to the plane of the sky, the SW component appears internally disturbed and may be undergoing a minor merger of its own, regardless of the chosen scenario.

Our competing hypotheses regarding the merger history of MACSJ0416 make clear and testable predictions for the X-ray surface brightness and ICM temperature distribution around the NE component and for the region between the two subclusters. Forthcoming deep \emph{Chandra} ACIS-I observations of MACSJ0416, capable of detecting the signature of sloshing gas and of temperature variations characteristic of ICM-ICM collisions thus hold great promise for a dramatically improved understanding of the formation history of this complex cluster lens.

\section*{Acknowledgments}

We thank St\'ephane Arnouts, Olivier Hilbert, and Christophe Adami for fruitful discussions. MJ thanks Ian Smail, and John Stott for their suggestions.
We thank Keiichi Umetsu and Daniel Gruen for sharing some of their results with us for comparison.
This work was supported by the Leverhulme Trust (grant number PLP-2011-003) and Science and Technology Facilities Council (grant number ST/L00075X/1). MJ, ML, and EJ acknowledge the M\'esocentre d'Aix-Marseille Universit\'e (project number: 14b030).  This study also benefited from the facilities offered by CeSAM (CEntre de donn\'eeS Astrophysique de Marseille ({\tt http://lam.oamp.fr/cesam/}). ML acknowledges the Centre National de la Recherche Scientifique (CNRS) for its support. JR acknowledges support from the ERC starting grant CALENDS and the CIG grant 294074. JPK and HA acknowledge support from the ERC advanced grant LIDA. DH is supported by a STFC studentship. PN acknowledges support from the National Science Foundation via the grant AST-1044455, AST-1044455, and a theory grant from the Space Telescope Science Institute HST-AR-12144.01-A. RM is supported by the Royal Society.
Based on observations made with the NASA/ESA Hubble Space Telescope, obtained from the data archive at the Space Telescope Science Institute. STScI is operated by the Association of Universities for Research in Astronomy, Inc. under NASA contract NAS 5-26555.

\bibliographystyle{mn2e}
\bibliography{reference}


\label{lastpage}

\end{document}